\documentclass[twocolumn,showpacs,preprintnumbers,amsmath,amssymb]{revtex4}
\usepackage{graphicx}
\usepackage{amsmath}

\begin{document}
\title{Localization and delocalization in the quantum kicked prime number rotator}
\author{Tao Ma}
\affiliation{Department of Modern Physics, University of Science and
Technology of China, Hefei, PRC}
\email{taomascience@gmail.com}
\date{\today}
\begin{abstract}
The quantum kicked prime number rotator (QKPR) is defined as the
rotator whose energy levels are prime numbers. The long time
behavior is decided by the kick period $\tau$ and kick strength $k$.
When $\frac{\tau}{2\pi}$ is irrational, QKPR is localized because of
the equidistribution theorem. When $\frac{\tau}{2\pi}$ is rational,
QKPR is localized for small $k$, because the system seems like a
generalized kicked dimer model. We argue for rational
$\frac{\tau}{2\pi}$ QKPR delocalizes for large $k$.
\end{abstract}
\maketitle

The kicked prime number rotator is defined as
\begin{equation}
H=H_0-V \sum_{n=1}^\infty \delta(t-n\tau),
\end{equation}
where $H_0$ is the unperturbed Hamiltonian, and $V$ is the
perturbation. $H_0$ is a diagonal matrix. The $m$-th eigenvalue
$E_m$ corresponding to $m$-th eigenstate $|m \rangle$ of $H_0$ is
the $|m|$-th prime number $p_{|m|}$. When $m<0$, $E_m=E_{-m}$.
$E_0=0$. The diagonal of $H_0$ is $\{
\ldots,11,7,5,3,2,0,2,3,5,7,11,\ldots \}$.

$V$ is defined as
\begin{equation}
V=\left(
\begin{array}{llllll}
 \cdots & \cdots &   &   &   &   \\
 \cdots & 0 & k/2 &   &   &   \\
   & k/2 & 0 & k/2 &   &   \\
   &   & k/2 & 0 & k/2 &   \\
   &   &   & k/2 & 0 & \cdots \\
   &   &   &   & \cdots & \cdots.
\end{array}
\right)
\end{equation}

The Floquet operator is
$F=e^{-\frac{i}{\hbar}V(\theta)}e^{-\frac{i}{\hbar} H_{0} \tau}$.
The matrix elements of $F$ is $F_{nm}=\exp(\frac{-i \tau E_m}{
\hbar}) i^{m-n} J_{n-m}(\frac{k}{\hbar})$, where $E_m=p_{|m|}$,
$J_{n-m}$ is the Bessel function of the first kind. We set
$\hbar=1$. $F_{nm}=\exp(-i \tau E_m) i^{m-n} J_{n-m}(k)$. The system
is very like the quantum kicked rotator (QKR), except its energy
levels are now prime numbers.

It seems there is no classical correspondence of QKPR.  Experimental
implementation of such a model also seems impossible. Nevertheless
it still has some theoretical interests. In the paper, we
numerically calculate the evolution of QKPR. We are interested in
the same problem in QKR. If the particle is in the ground state $|0
\rangle$ initially, will it diffuse away in the future?

The evolution of the system is calculated by the iterative unitary
matrix multiply method \cite{TaoMa2007}. $F^4=(F^2)^2$.
$F^4=(F^2)^2$. $F^8=(F^4)^2$. And so on. In this way, we can
calculate $F^{(2^{50})}$ by $50$ matrix multiplies. In all our
calculation, $N$ indicates at time $2^N\tau$. For example, the first
figure $N=10$ means at time $2^{10}\tau=1024\tau$. $n$ is the $n$-th
basis $|n \rangle$ and $c_n$ is the base-10 logarithm of the
absolute value of the wave function on the $|n \rangle$. $n$ runs
from $-500$ to $500$ in our calculation.

First, we choose $k=1$, $\tau=2 \pi \frac{\sqrt{5}-1}{2}$. The
result is displayed in FIG. 1. QKPR is localized perfectly. In our
simulation, the exponentially fall of the wave function never
changes from $N=8$ to $N=50$. The wave function on the $|n \rangle$
is $10^{c_n}$. From $N=1$ to $N=7$, the wave function is somewhat
curved. After the first kick, the wave function is the $0$-th column
of the Floquet matrix $F$. The absolute value of $F_{n0}$ is
$|J_n(k)|$. $|J_n(k)|$ falls to zero faster than exponentially .
This is the reason the curved form of the wave function.

Second, we choose $k=5$, $\tau=2 \pi \frac{\sqrt{5}-1}{2}$. The
result is displayed in FIG. 3. The wave function is also localized.
This is expected. The sequence $\{-p_n \frac{\tau}{2\pi} Mod 1\}$ is
equidistributed between $[0,1]$, when $\frac{\tau}{2\pi}$ is
irrational. We denote the sequence $\{-p_n \frac{\sqrt{5}-1}{2} Mod
1\}$ as QKPR$_G$. In QKR, the sequence $\{-\frac{n^2}{2}
\frac{\tau}{2\pi} Mod 1\}$ is also equidistributed between $[0,1]$,
for an irrational $\frac{\tau}{2\pi}$. We denote the sequence
$\{-\frac{n^2}{2} \frac{\sqrt{5}-1}{2} Mod 1\}$ as (QKR$_G$). We can
also use the inverse Cayley transform method to convert the Floquet
eigenstate equation $F \varphi=\lambda \varphi$ into an equation
like Anderson localization problem. From Fishman \textit{et al}'s
argument \cite{Fishman1982}, QKPR will localize.

In the left of FIG. 2, QKPR$_G$ and QKR$_G$ are displayed. Though
there are apparently some correlations in QKPR$_G$ and QKR$_G$ and
the correlation is different between both sequences. The correlation
is surely not strong enough to destroy localization. If a sequence
is periodic with a period $q$, then the discrete Fourier transform
of the sequence is composed by $q$ modes. To find whether there is
some periodicity in the sequence, we perform a discrete Fourier
transform (DFT) on the sequence. DFT of a sequence $s_n$ of length
$L$ is defined as $F_j=\sum_{n=1}^{L}s_n e^{-i 2\pi (n-1)(j-1)/L}$,
where $j$ runs from 1 to $L$. There are some other definitions of
DFT with nuanced difference with our definition. But the difference
is irrelevant to our discussion here. In the right of FIG. 2, the
DFTs of both sequences are displayed. There are no rigorous
periodicity in both sequences. $F_k$ of QKPR$_G$ seems to have a
trend to cluster together. Also it is less uniformly distributed
than the $F_j$ of QKR$_G$ and tends to be small.

If $\frac{\tau}{2\pi}=\frac{1}{3}$ , does QKPR localize? At first
thought, this seems to be a resonant case in QKR and the rotator
will delocalize. The calculation result is in fact it still
localizes for small $k$. In FIG. 4, we choose $k=1$ and $\tau=2 \pi
\frac{1}{3}$ and in FIG. 5, $k=5$ and $\tau=2 \pi \frac{1}{3}$. QKPR
of $k=1$ is apparently localized.

The explanation of the localization is QKPR with
$\frac{\tau}{2\pi}=\frac{1}{3}$ and $k=1$ is a kicked pseudo dimer
rotator. The dimer model is defined as every diagonal matrix element
is a probability variable which only takes two values
\cite{Bovier1991,Dunlap1989}. The kicked dimer model can be defined
as every diagonal matrix element of $e^{-iH_0\tau}$ is a random
variable which takes two values. If it takes more than two values,
it is a generalized kicked dimer model. For $q=5$, the sequence
$-p_n \frac{1}{q} Mod 1$ mainly takes four values. So it is a
generalized kicked pseudo dimer model. The sequence $-p_n \tau Mod
1$ (QKPR$_3$) is not really random. But the pseudorandomness is
enough to result in localization \cite{Fishman1982}. To measure how
random QKPR$_3$ is, we perform a DFT on it. In the left of FIG. 4,
we compare QKPR$_3$ with a dimer sequence D$_3$, which is defined as
$-\frac{1}{3}Random(n) Mod 1$, where every $Random(n)$ is a random
variable which takes two values $1$ and $2$. The $F_k$s of D$_3$ and
QKPR$_3$ are quite close with each other, except QKPR$_3$ tends to
cluster together.

Does QKPR localize for $\frac{\tau}{2\pi}=\frac{1}{3}$ and $k=5$? We
think it delocalizes. There are a series of plateaux in the wave
function of QKPR. The wave function falls abruptly when approaching
the boundary (cliff) of a plateau. Some plateaux disappear at $N=14,
27, 45$. QKPR wave pass through the cliff, so it disappears
intermittently.

The most obvious cliff is from $n=140$ to $160$ in FIG.6. QKPR$_3$
from $n=139$ to $161$ is
\begin{equation}
\begin{split}
\{&\frac{1}{3},\frac{1}{3},\frac{2}{3},\frac{1}{3},\frac{2}{3},
\frac{1}{3},\frac{2}{3},\frac{1}{3},\frac{2}{3},\frac{1}{3},\frac{2}{3},
\frac{1}{3},\\
&\frac{2}{3},\frac{1}{3},\frac{2}{3},\frac{1}{3},\frac{2}{3},
\frac{1}{3},\frac{2}{3},\frac{1}{3},\frac{2}{3},\frac{1}{3},\frac{1}{3}\}.
\end{split}
\end{equation}
Note it is periodic from $n=140$ to $160$. In a perfect periodic
potential the wave will always diffuse away. The quantum wave can
not stay at a period potential very long. Once the wave has
propagated into the phase space between $n=140$ and $160$, it
diffuses away quickly. While once the wave propagate the phase space
whose neighborhood $-p_n/3 Mod 1$ is irregular, the quantum wave is
localized there at least temporarily. When $k=1$, $n$ from $140$ to
$160$ is a plateau in FIG. 2. There is a transition from a plateau
to a cliff when $k$ becomes large.

Note in FIG. 4, for lots of $k$, Lots of $F_k$ of QKPR$_3$ and
QKPR$_G$ tend to be small. From Plancherel's theorem, there must be
some $F_k$ tends to very large. So there is some weak periodicity in
the sequence QKPR$_3$. The sequence from $n=140$ to $160$ increases
the periodicity of QKPR$_3$.

At $N=54$, there is a plateau from $-50$ to $50$ or so. Even at
$N=50$, there is a plateau from $-50$ to $50$. The localization
length of QKR is $\frac{k^2}{4}=6.25$ or so \cite{Chirikov1981,
Chirikov1988, Shepelyansky1986}. So the distribution length of QKPR
is much larger than the localization length. Even the result of
$N=54$ is untrustworthy, we think the quantum wave is absolutely not
localized in the localization length. From $n=-150$ to $150$ or so,
there is apparently unneglectable quantum wave at every $n$ at for
example $N=40$. At $N=6$, the quantum wave has already propagated
into $n \gg 6\frac{1}{4}$. QKPR with $\frac{\tau}{3}=\frac{1}{3}$
and $k=5$ is not localized. In \cite{TaoMa2007General}, we point out
when the kick strength is larger than $\pi$, the inverse Cayley
transform method breaks down. QKPR is apparently an evidence to the
failure of the inverse Cayley transform method when $k$ is large.

In this paper, we apply the iterative unitary matrix multiply method
to quantum kicked prime number rotator. If $\frac{\tau}{2\pi}$ is
irrational, the rotator localizes. If $\frac{\tau}{2\pi}$ is
rational, for small kick strength $k$, the rotator localizes. As $k$
increases, we argue there is a localization-delocalization
transition.
\begin{figure*}
\begin{center}
   \begin{minipage}{17.0 cm}
   \includegraphics[width=5.5cm]{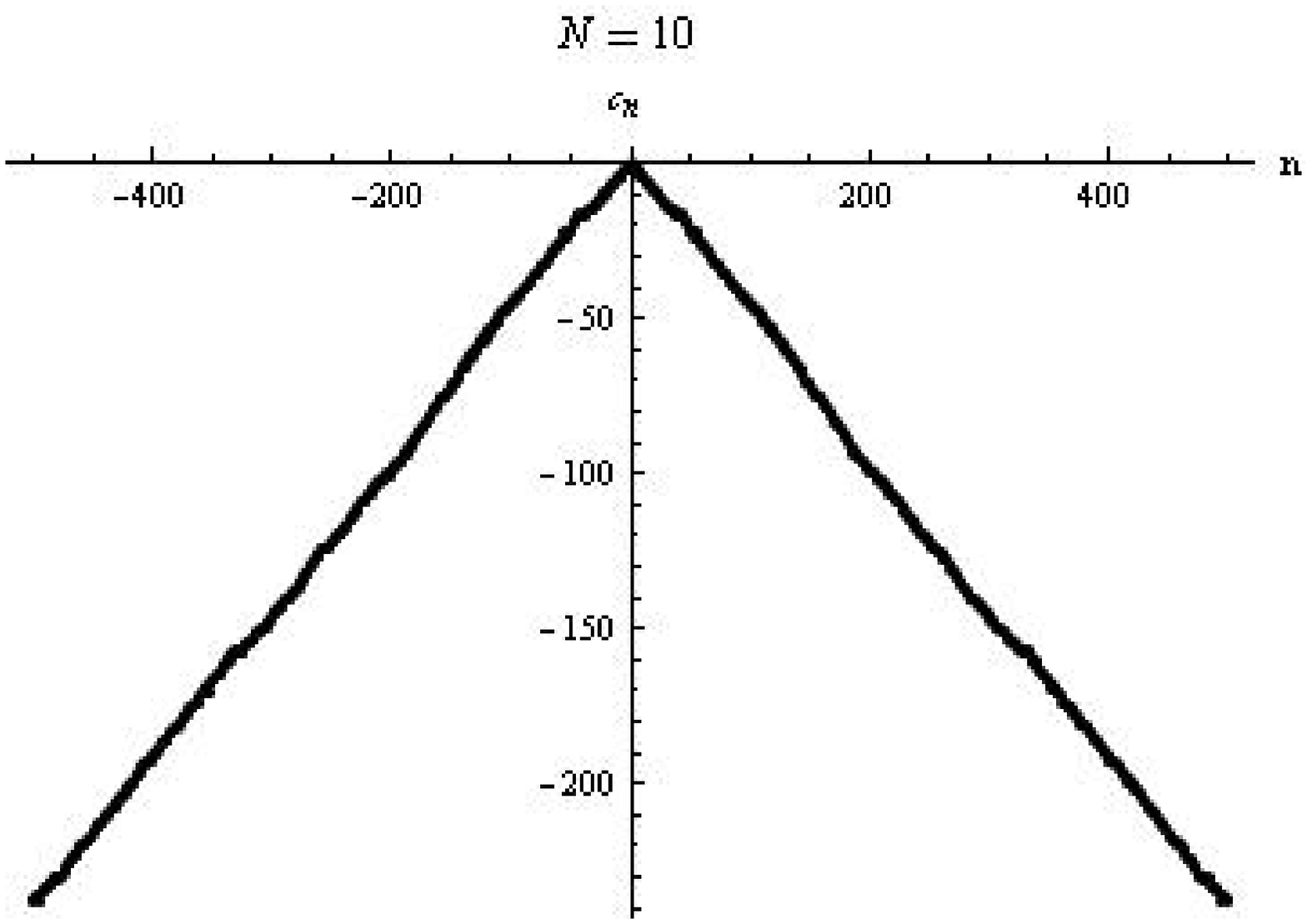}
   \includegraphics[width=5.5cm]{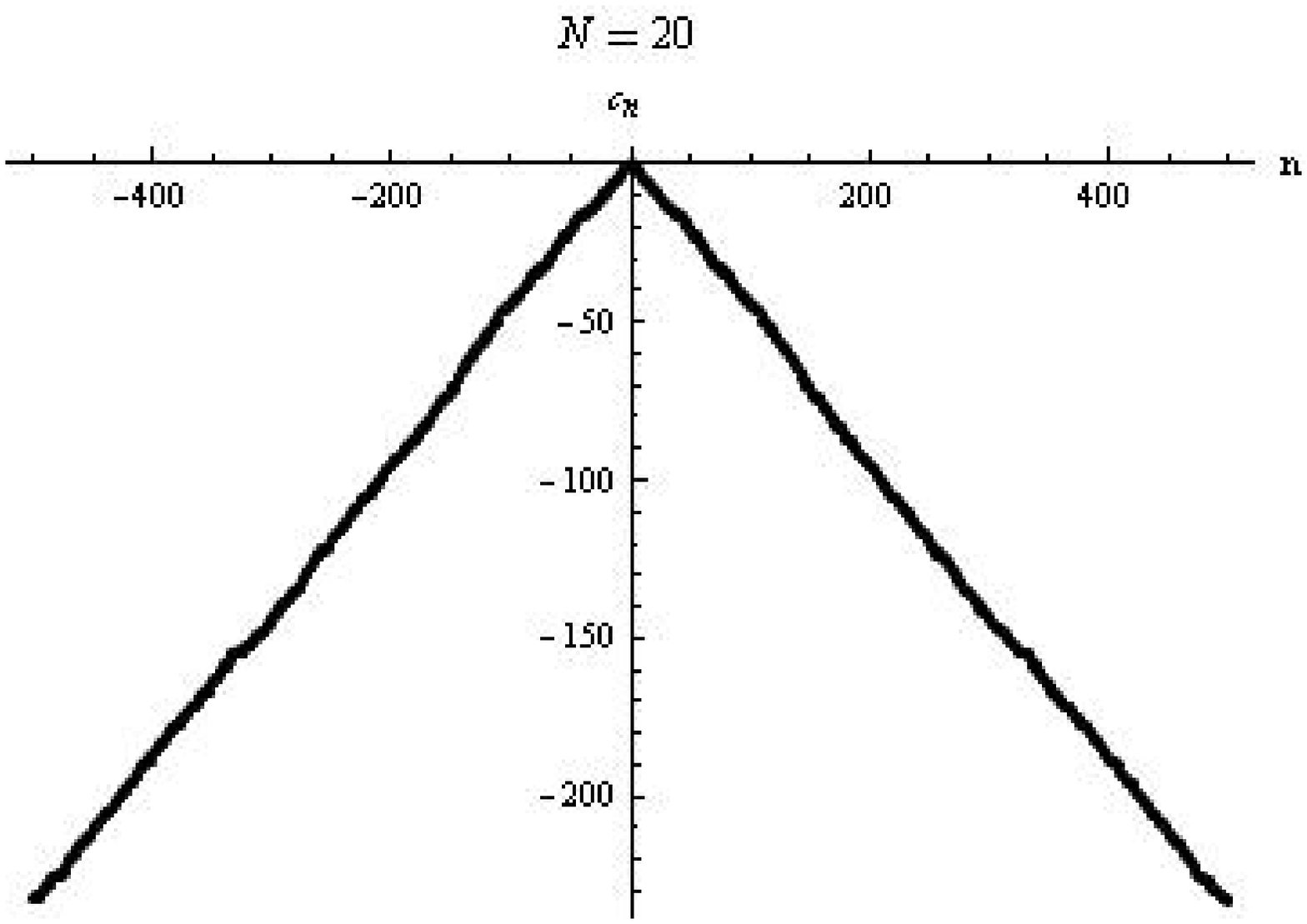}
   \includegraphics[width=5.5cm]{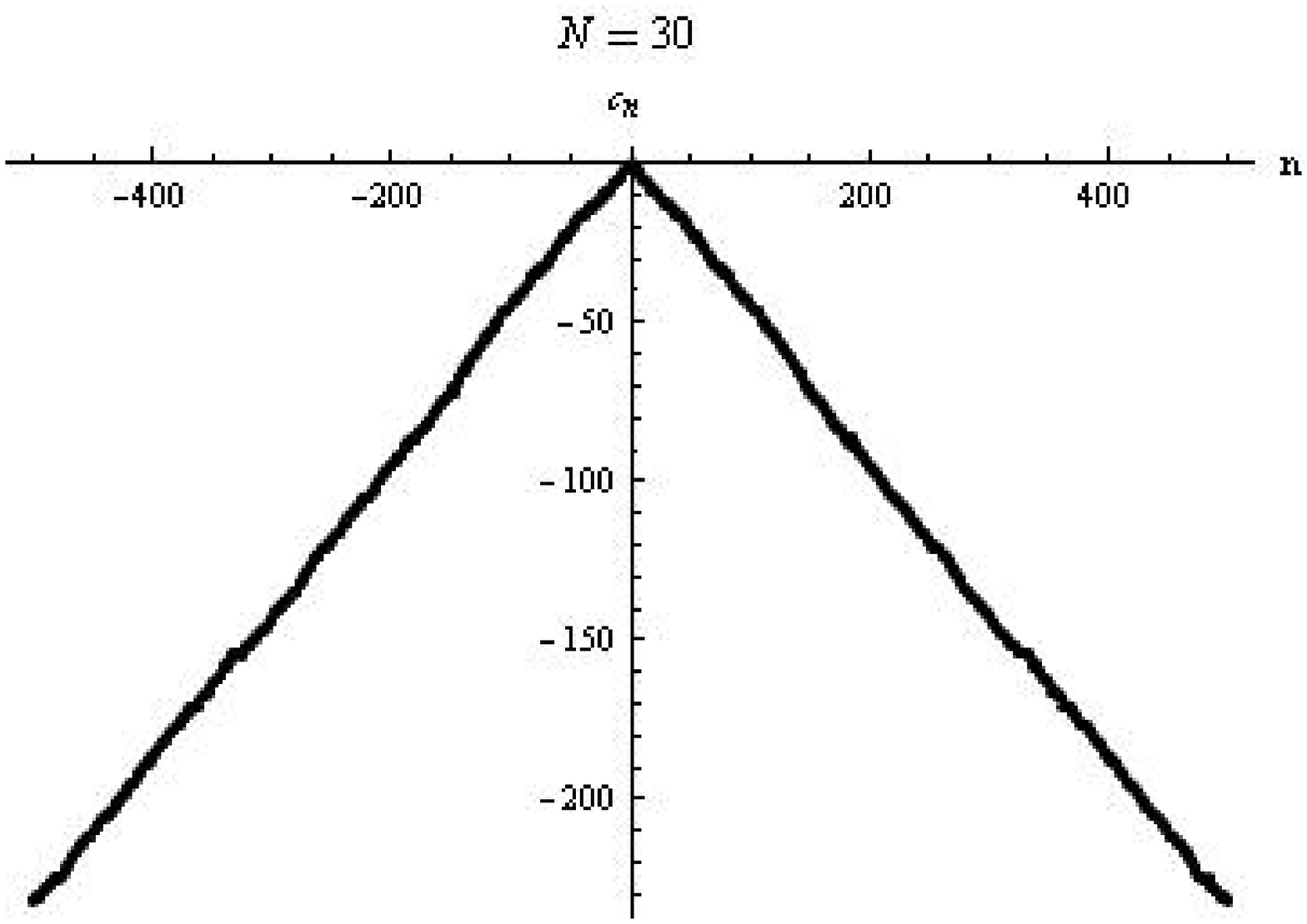}
\caption{QKPR wave function at different time for $k=1$, $\tau=2 \pi
\frac{\sqrt{5}-1}{2}$.}
\end{minipage}
   \begin{minipage}{17.0 cm}
   \includegraphics[width=6.0cm]{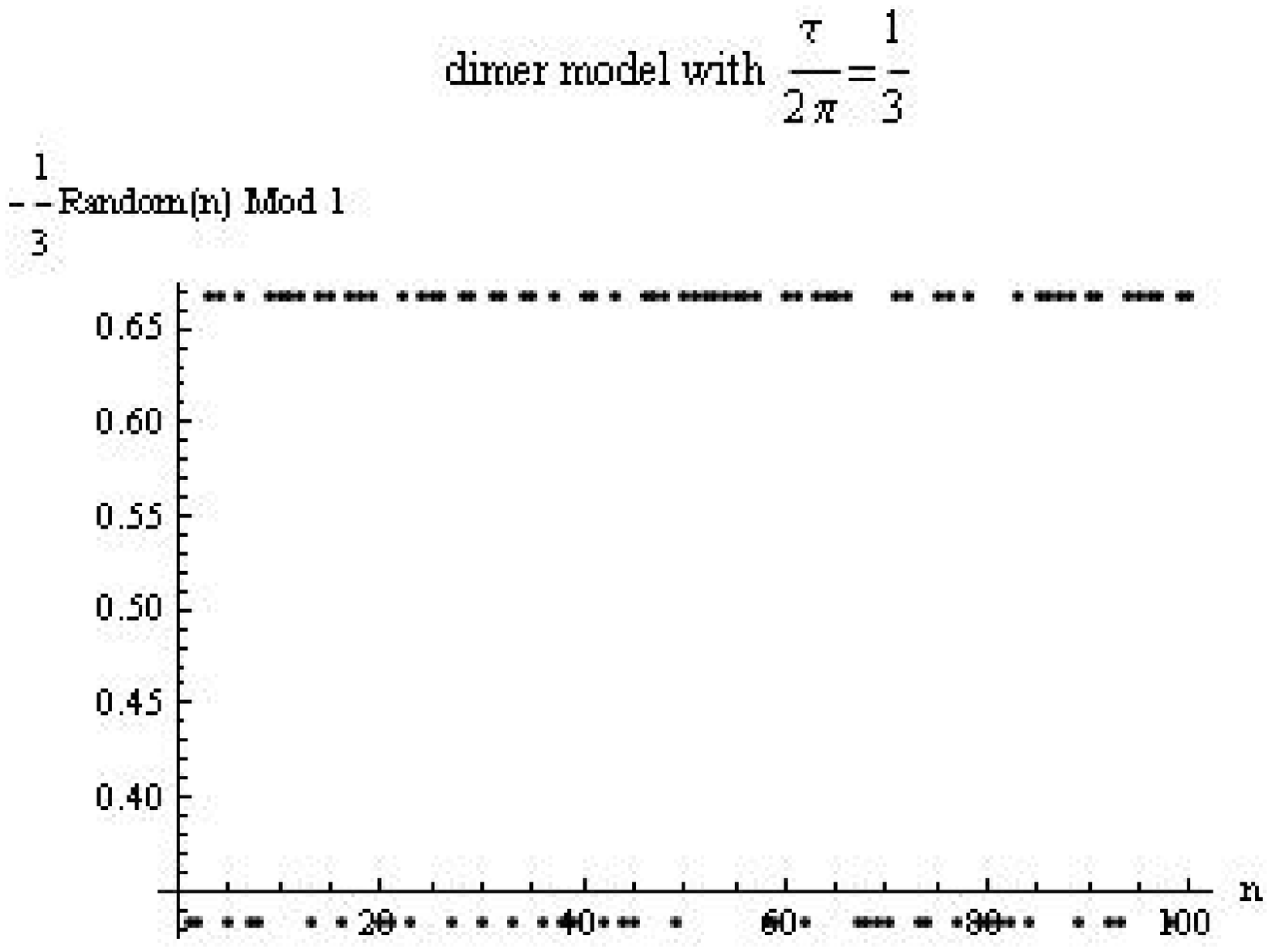}
   \includegraphics[width=6.0cm]{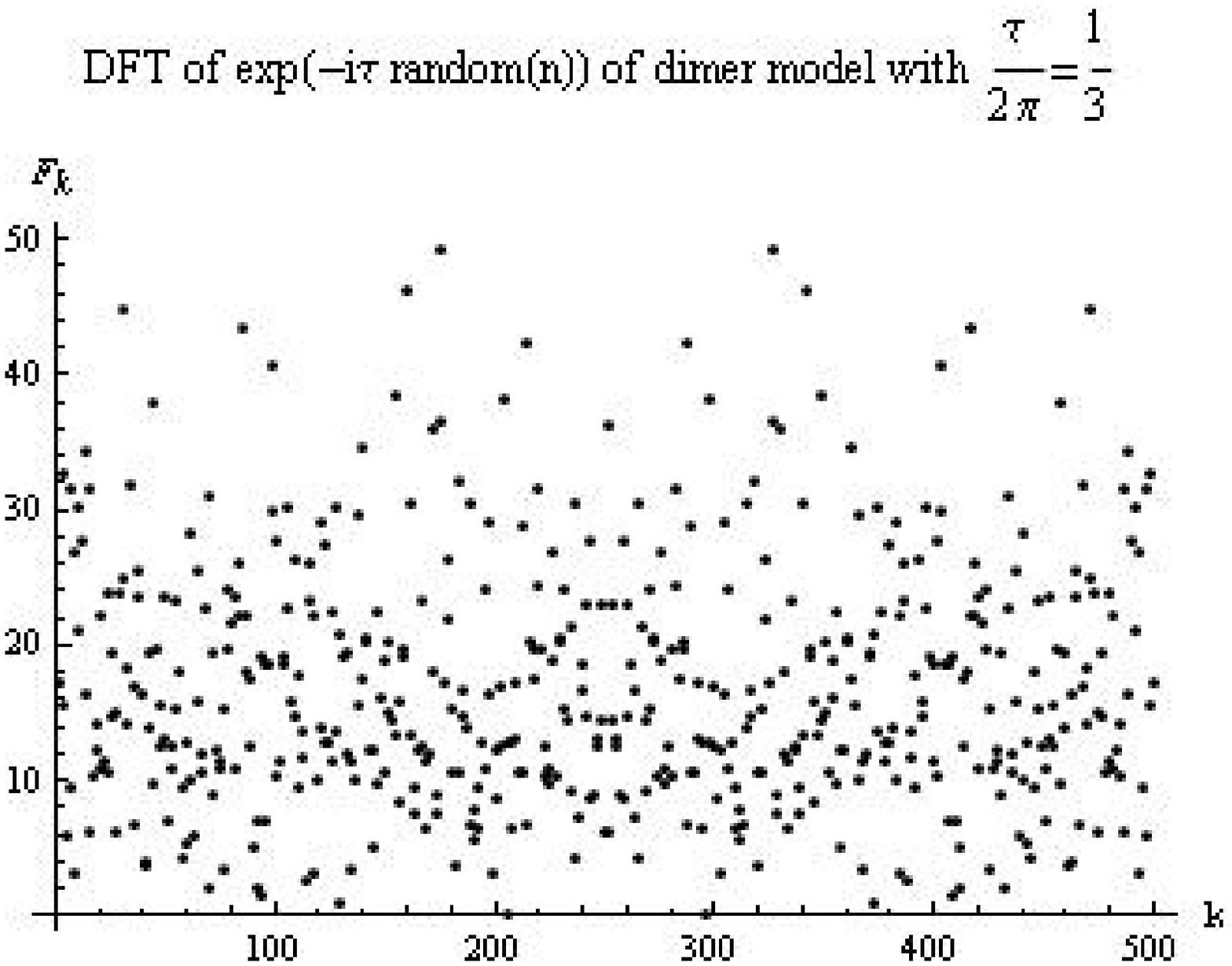}
  \end{minipage}
   \begin{minipage}{17.0 cm}
   \includegraphics[width=6.0cm]{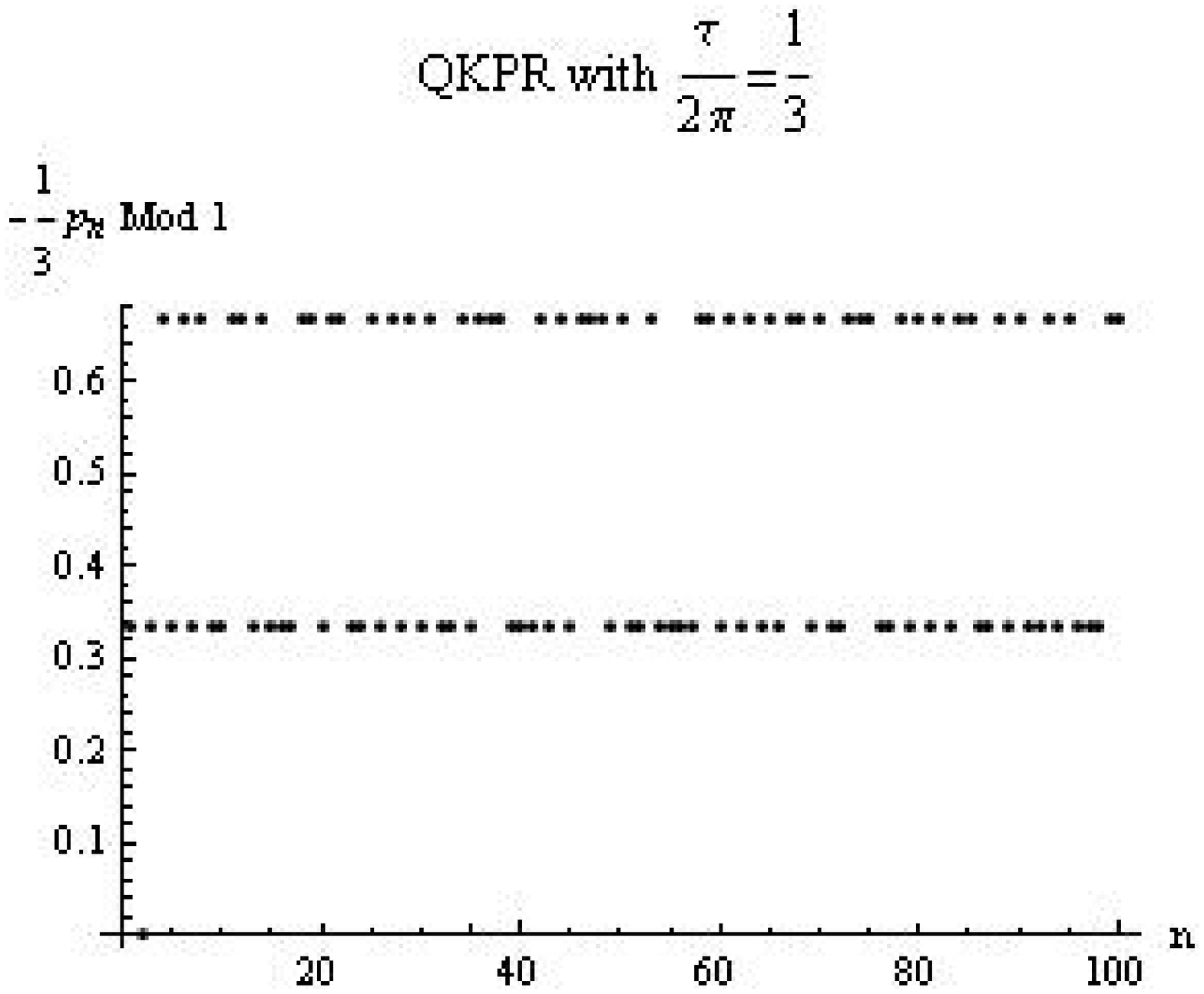}
   \includegraphics[width=6.0cm]{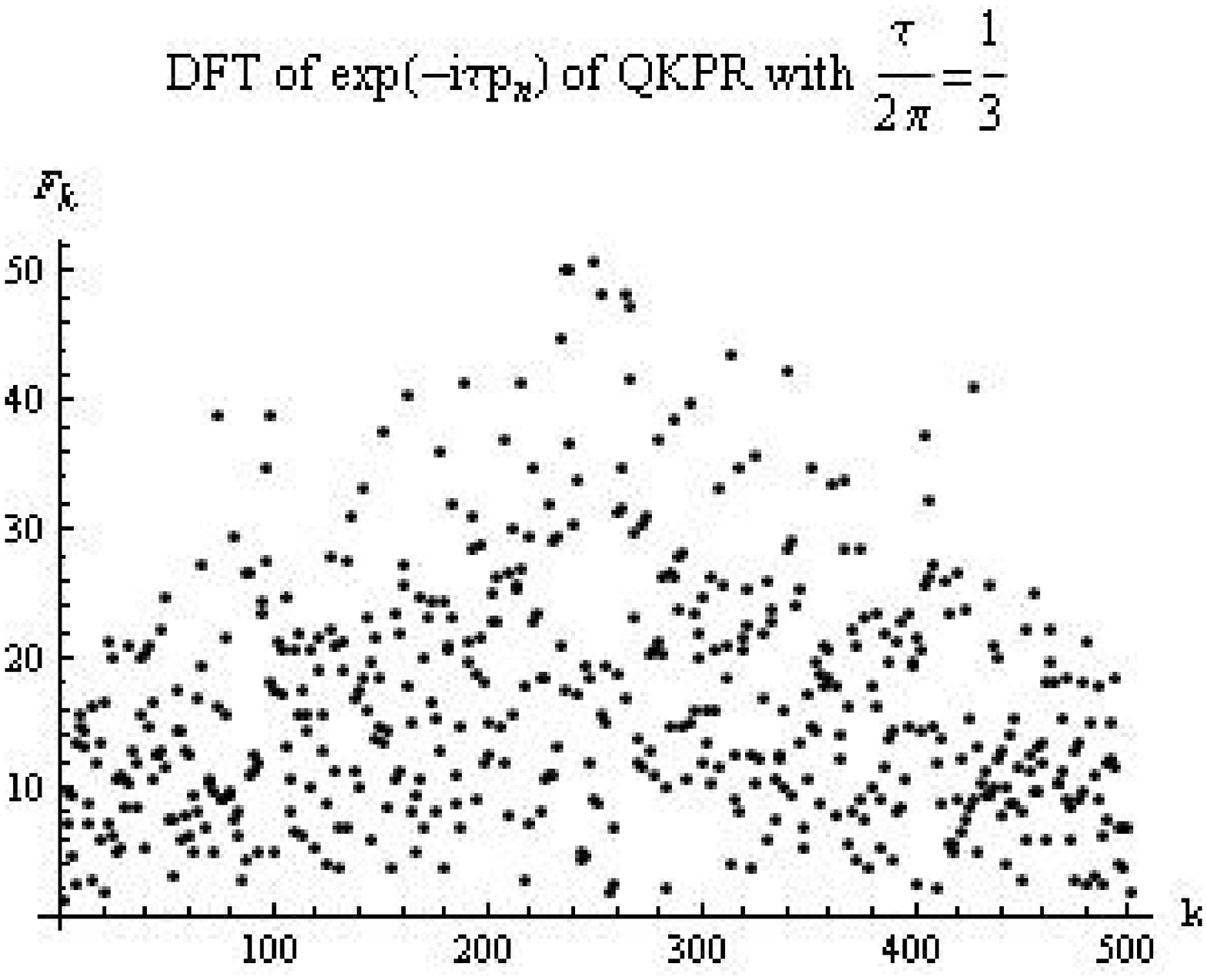}
  \end{minipage}
   \begin{minipage}{17.0 cm}
   \includegraphics[width=6.0cm]{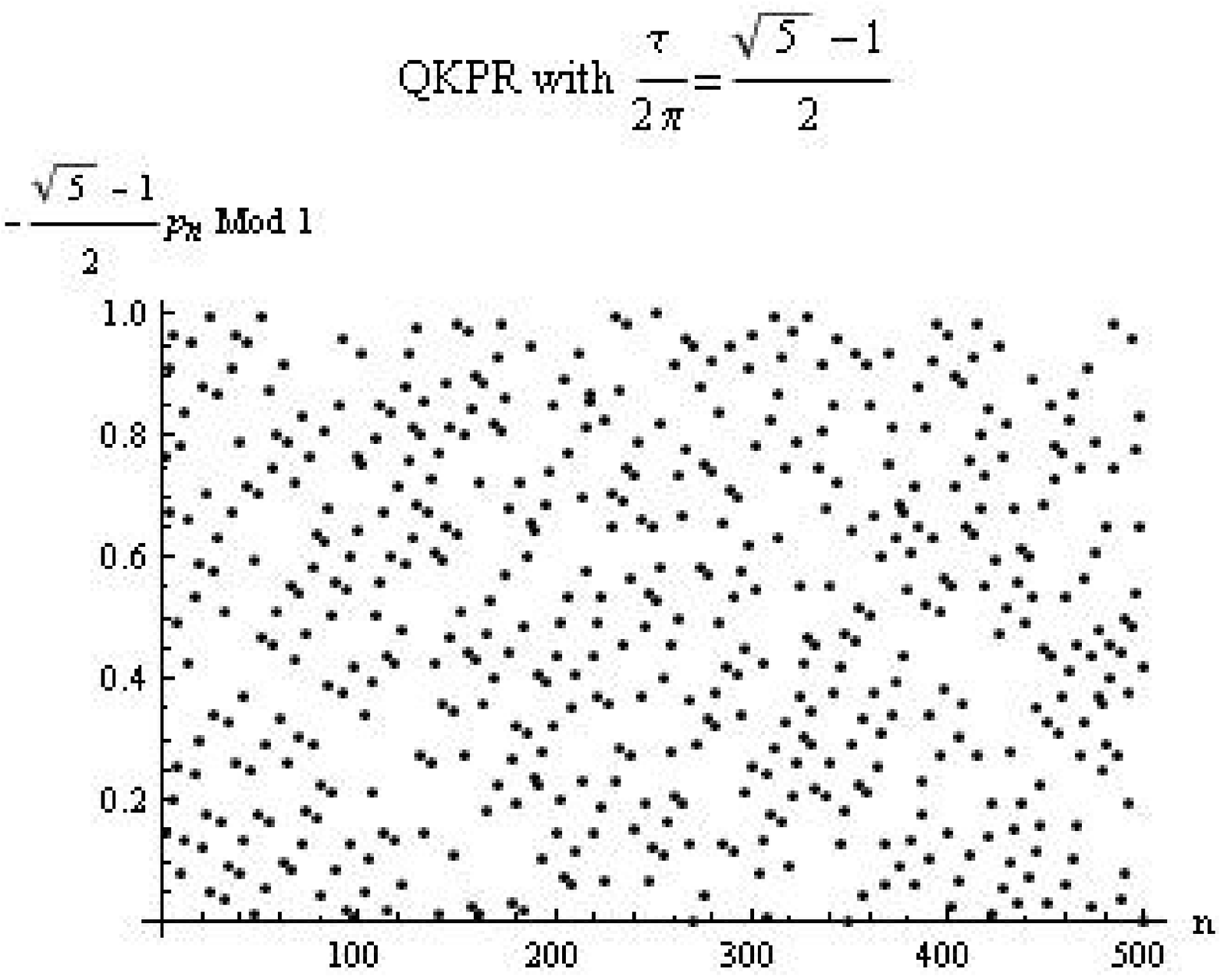}
   \includegraphics[width=6.0cm]{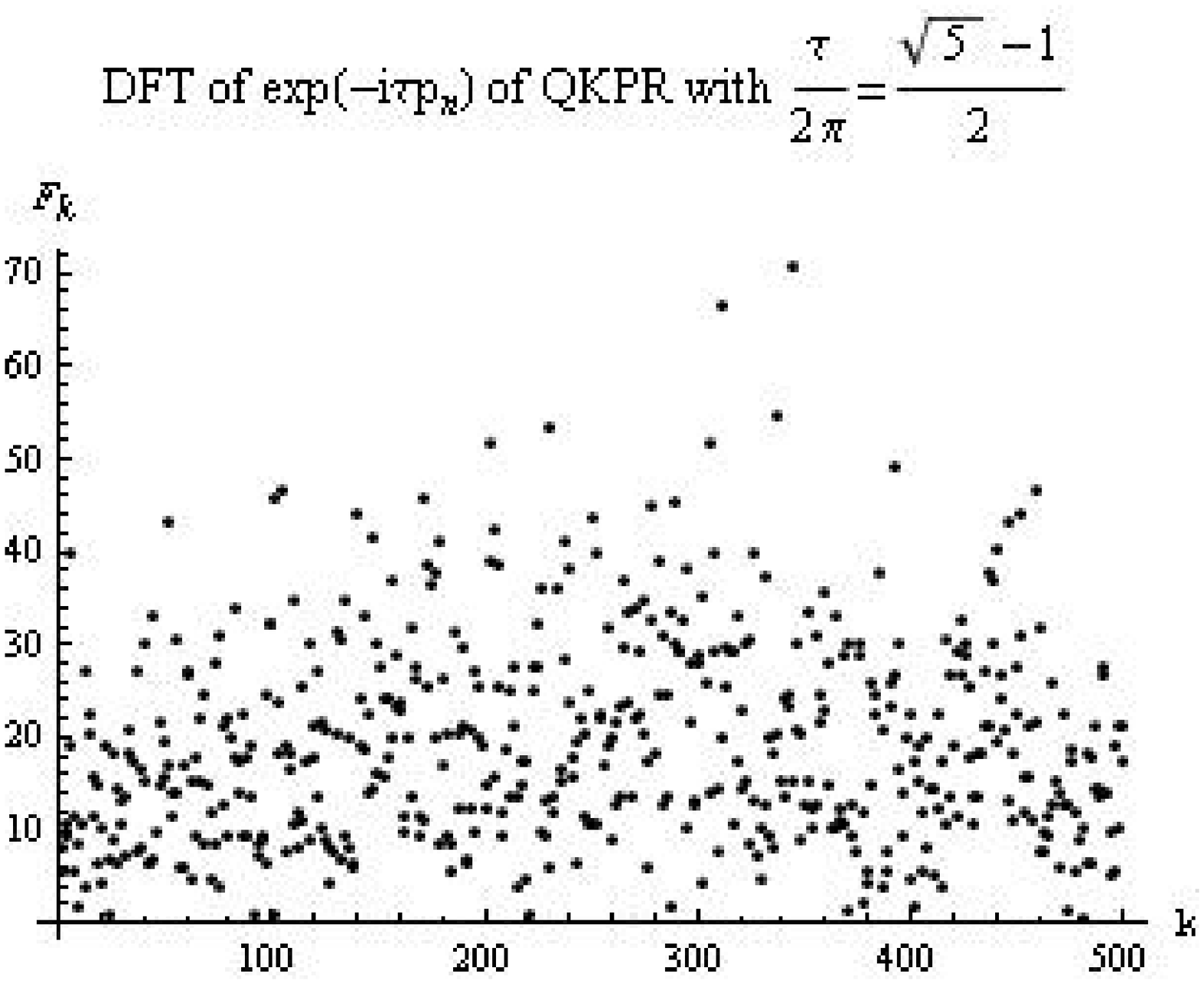}
   \end{minipage}
    \begin{minipage}{17.0 cm}
   \includegraphics[width=6.0cm]{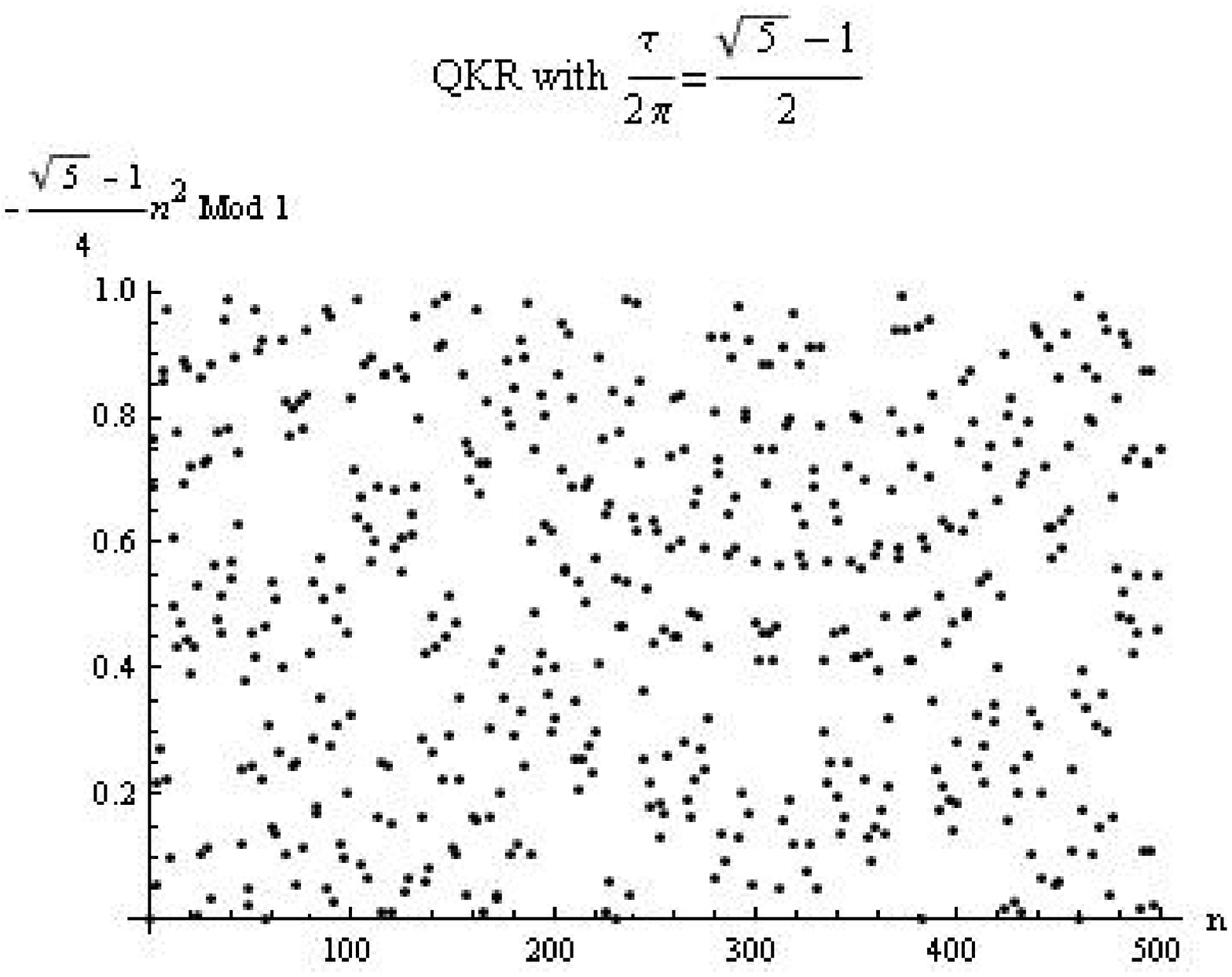}
   \includegraphics[width=6.0cm]{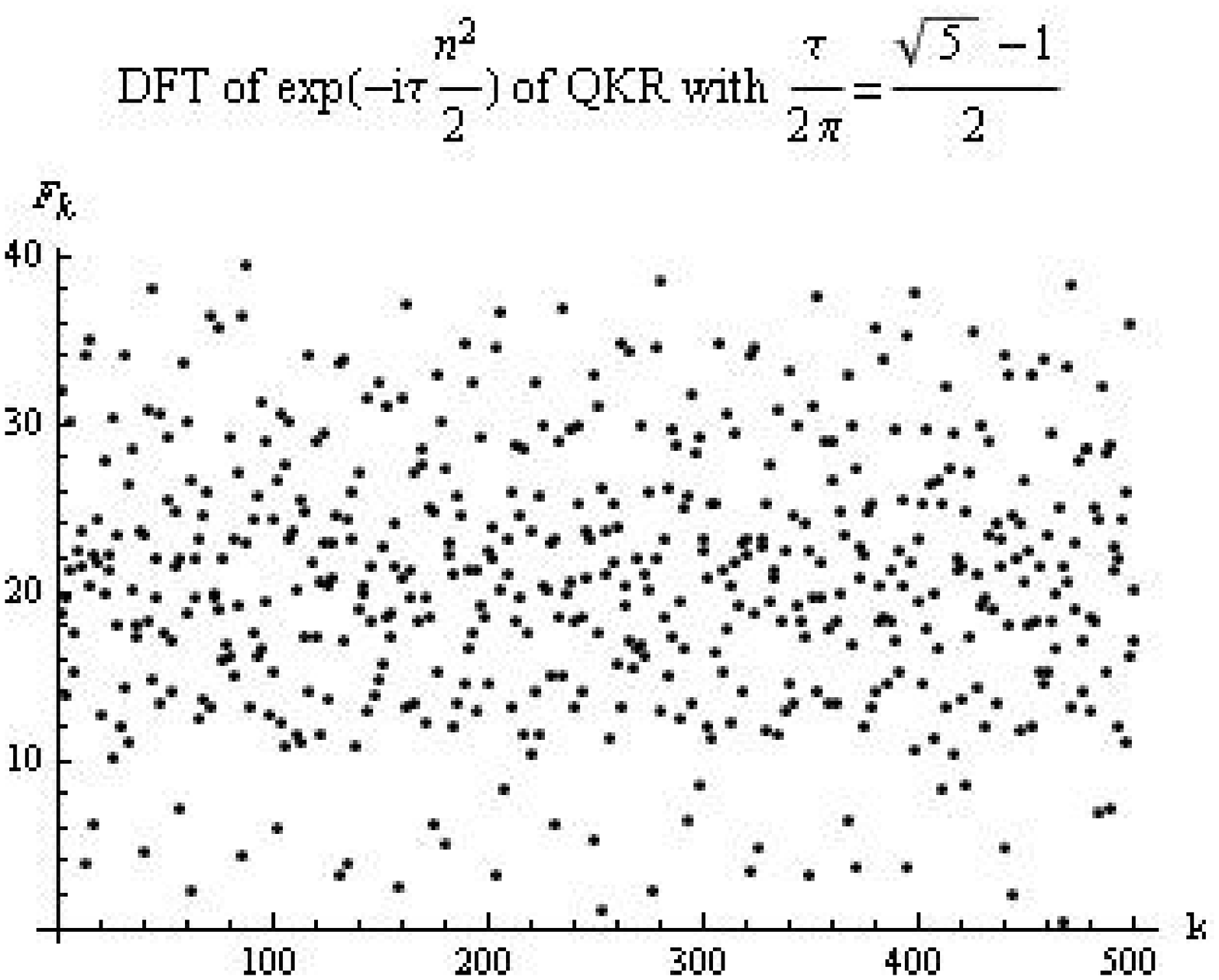}
\caption{Left: $-E_n \tau$ modulo 1, where $E_n$ is the $n$-th
energy level. Right: DFT of the sequence $\{ exp(-i E_n \tau) \}$,
where $n$ runs from 1 to 500.}
    \end{minipage}
\end{center}
\end{figure*}

\begin{figure*}
\begin{center}
   \begin{minipage}{17.0 cm}
   \includegraphics[width=5.5cm]{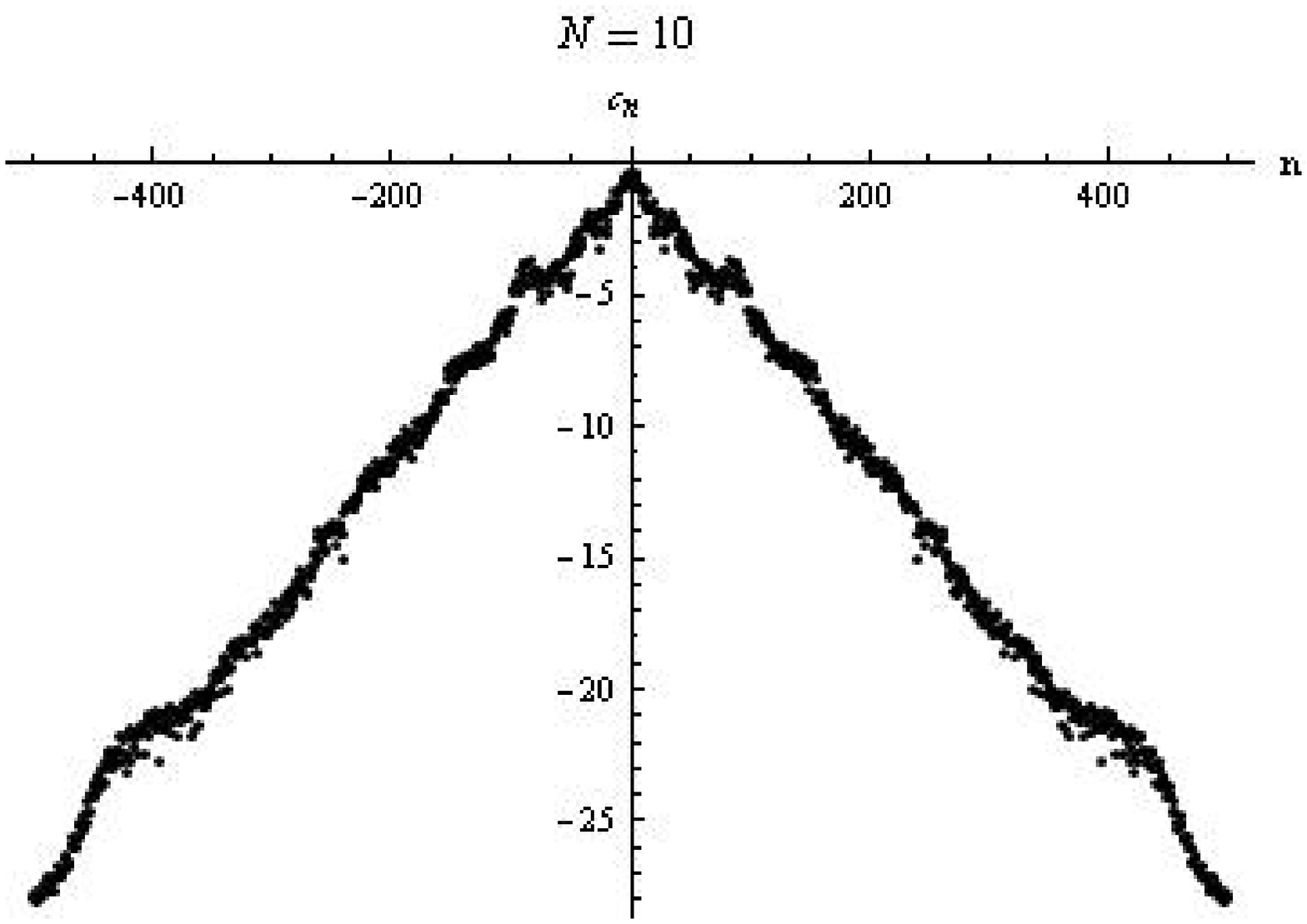}
   \includegraphics[width=5.5cm]{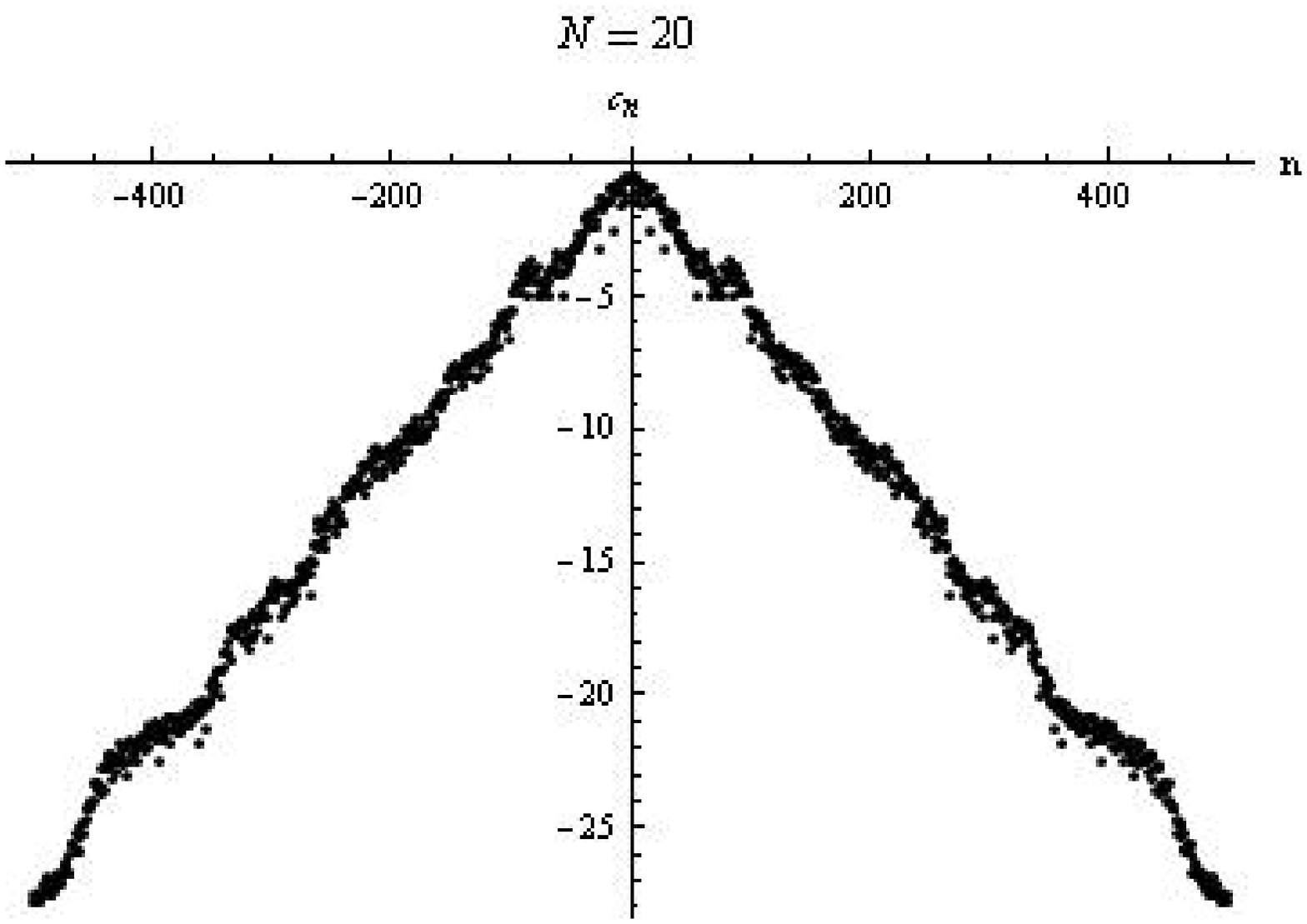}
   \includegraphics[width=5.5cm]{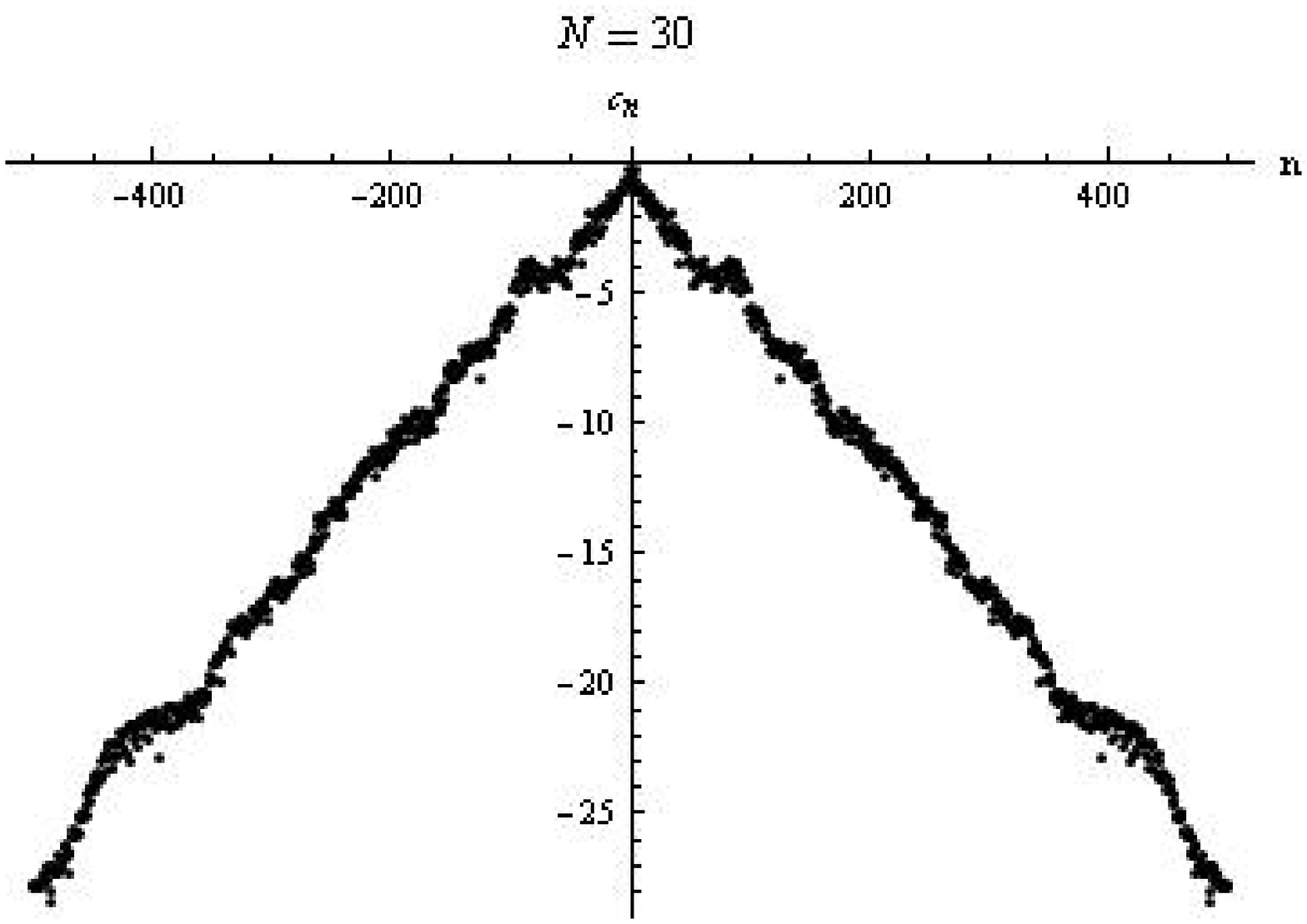}
\caption{QKPR wave function at different time for $k=5$, $\tau=2 \pi
\frac{\sqrt{5}-1}{2}$.}
\end{minipage}
   \begin{minipage}{17.0 cm}
   \includegraphics[width=5.5cm]{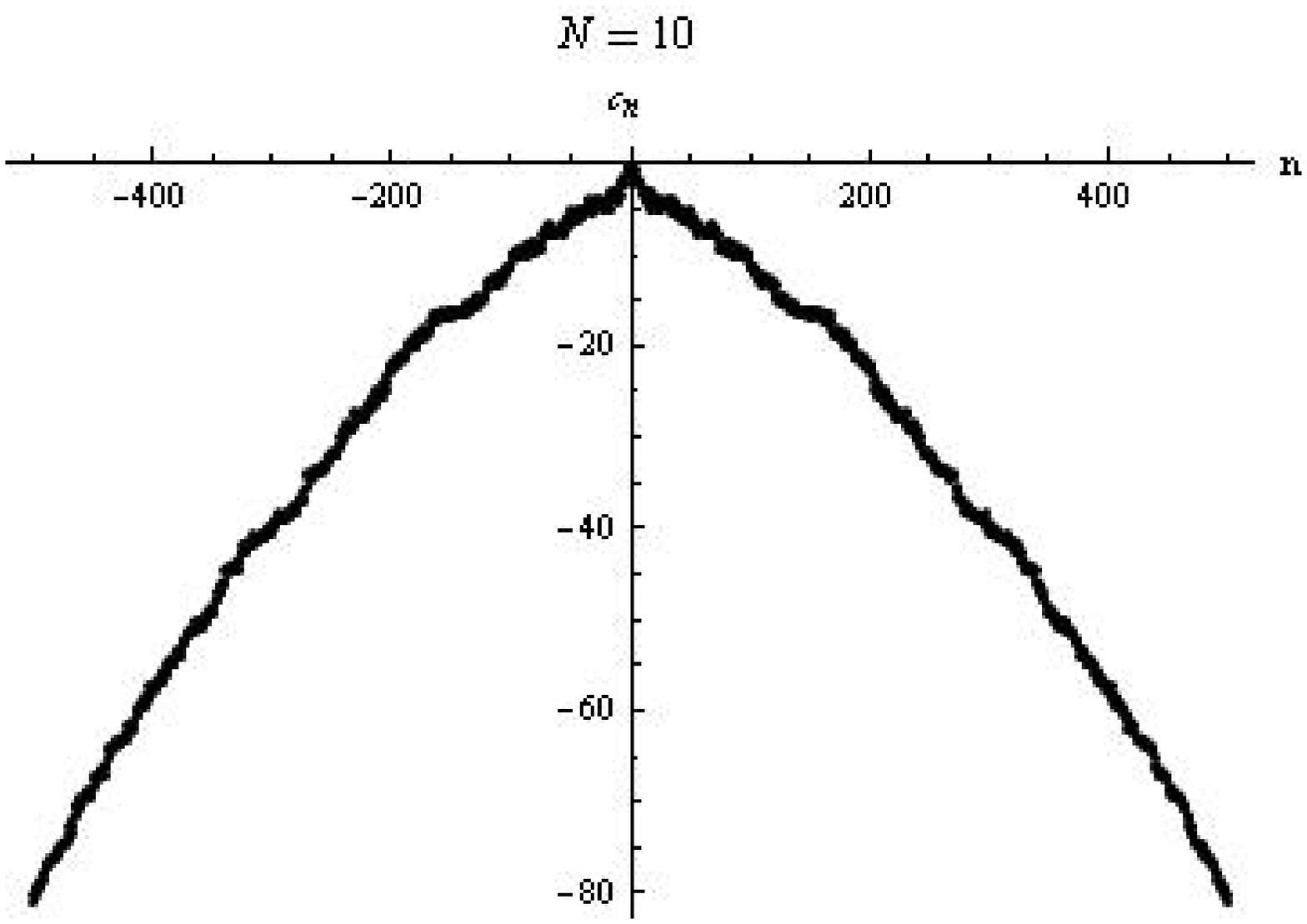}
   \includegraphics[width=5.5cm]{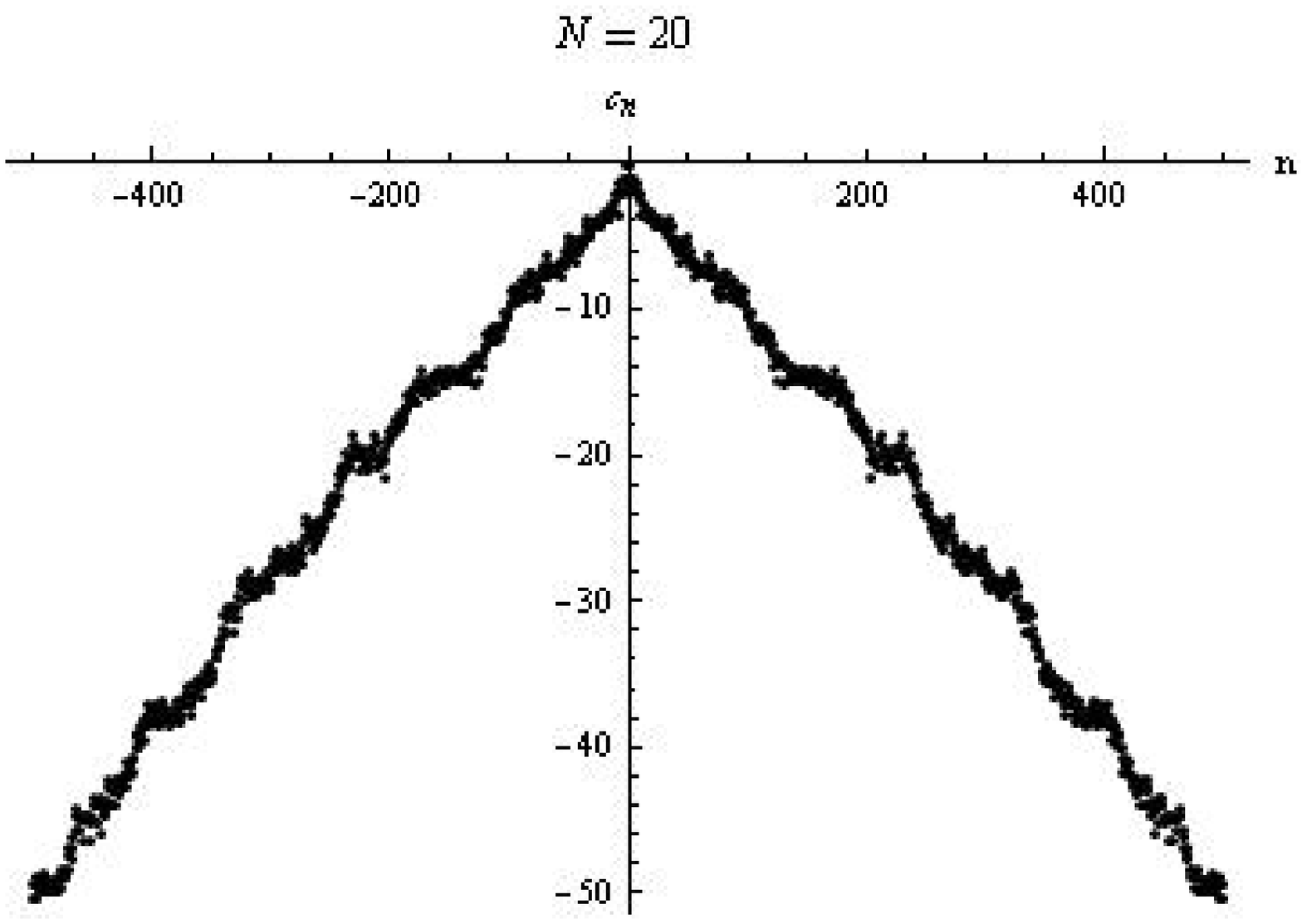}
   \includegraphics[width=5.5cm]{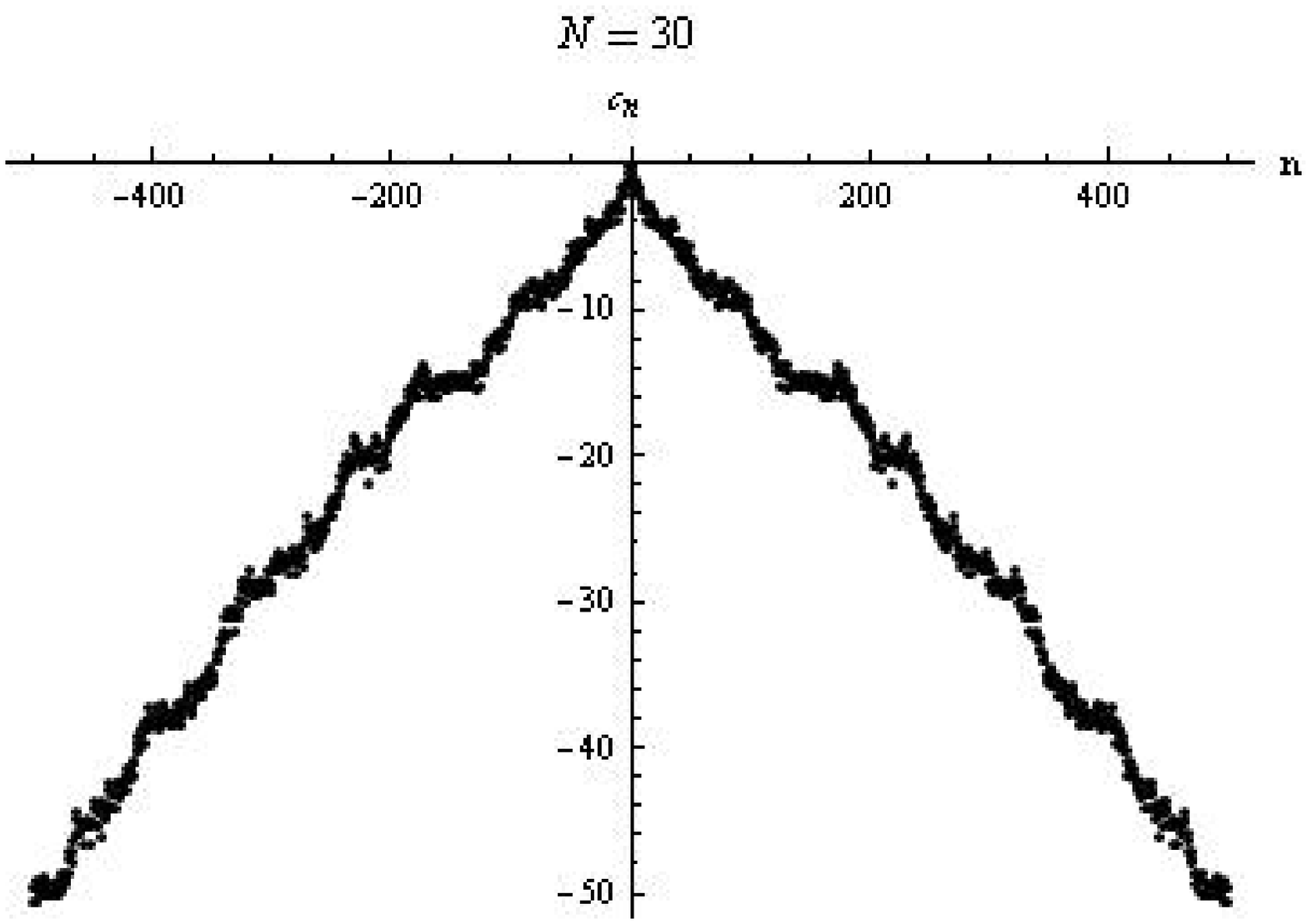}
\caption{QKPR wave function at different time for $k=1$, $\tau=2 \pi
\frac{1}{3}$.}
\end{minipage}
   \begin{minipage}{17.0 cm}
   \includegraphics[width=5.5cm]{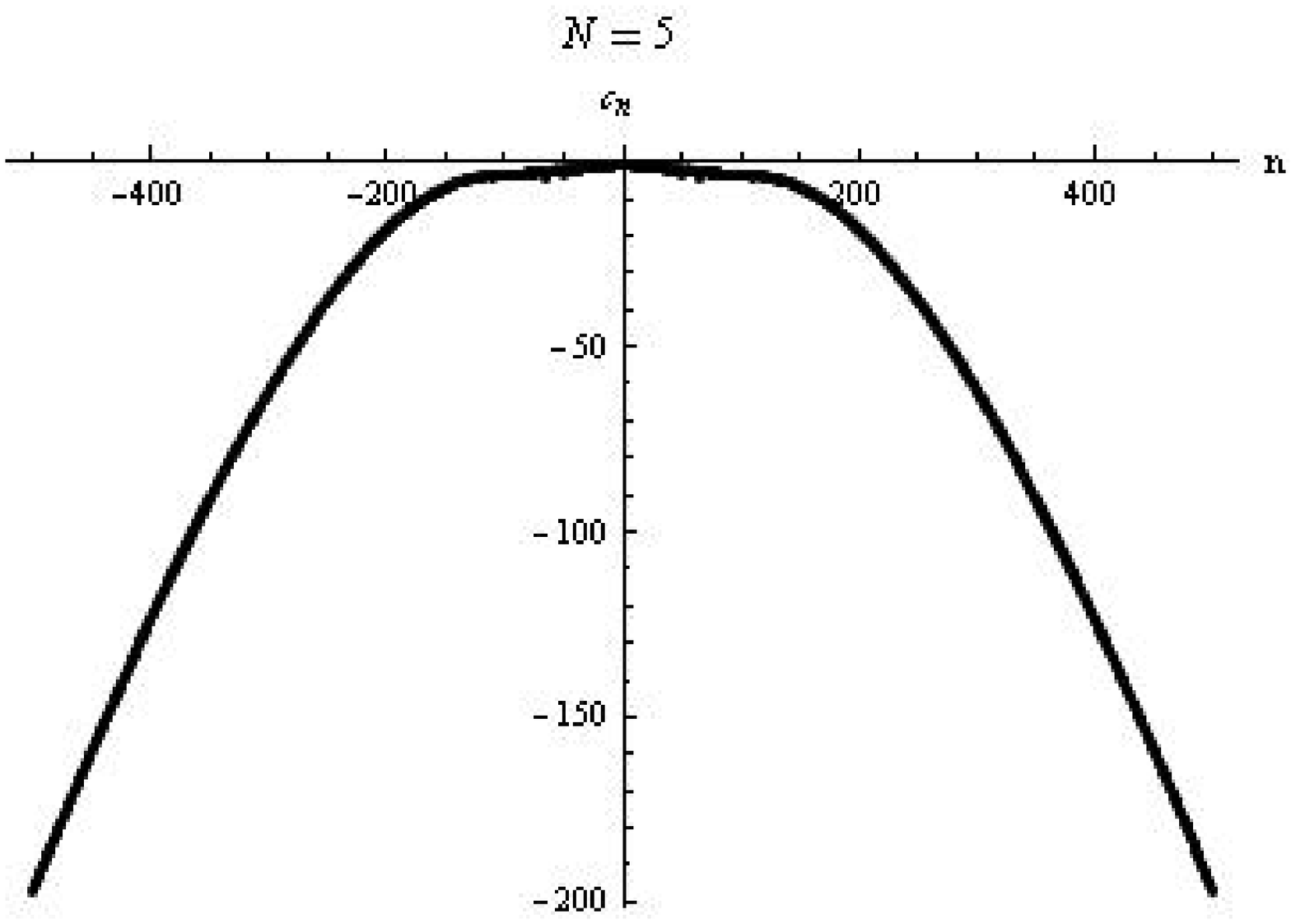}
   \includegraphics[width=5.5cm]{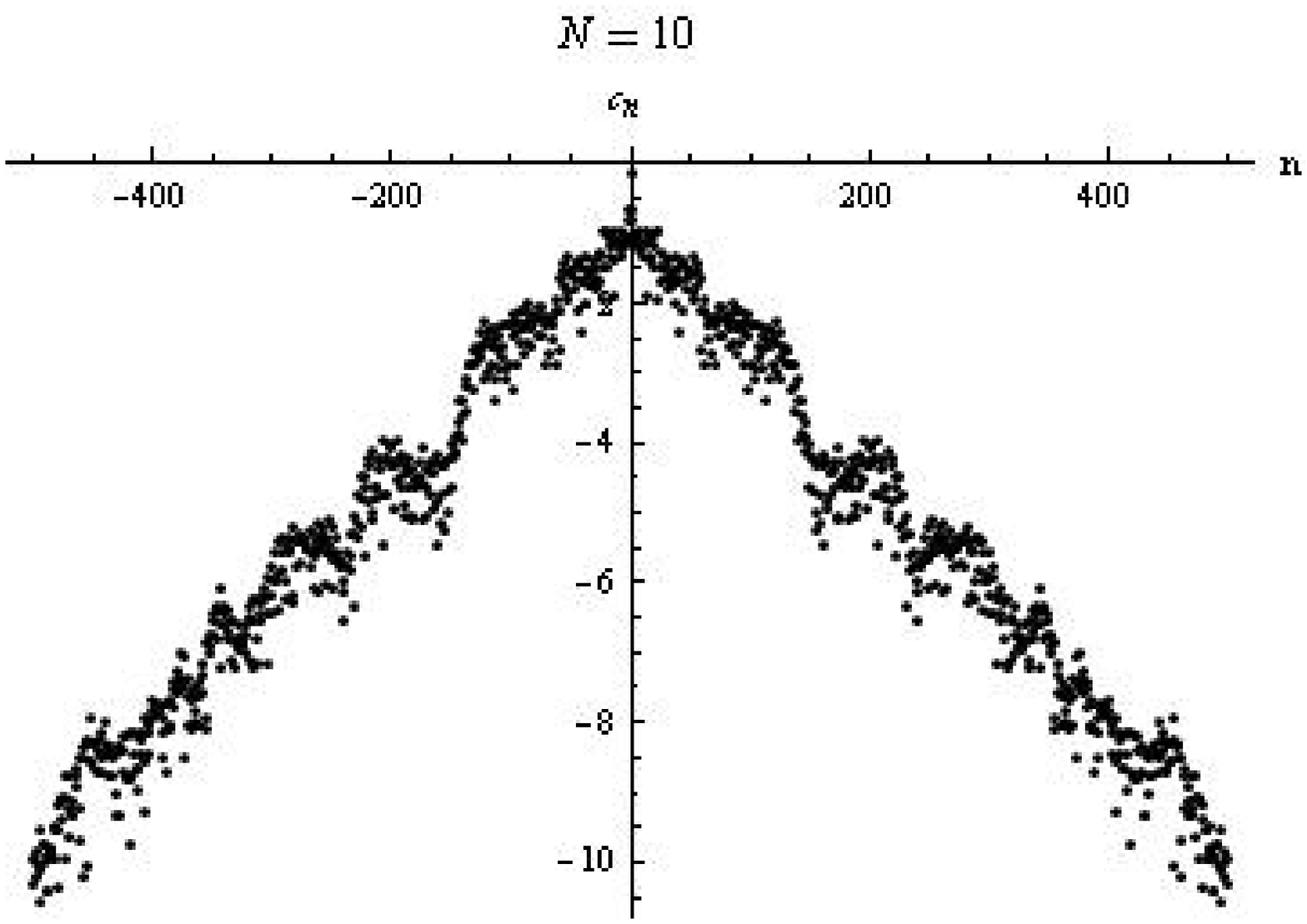}
   \includegraphics[width=5.5cm]{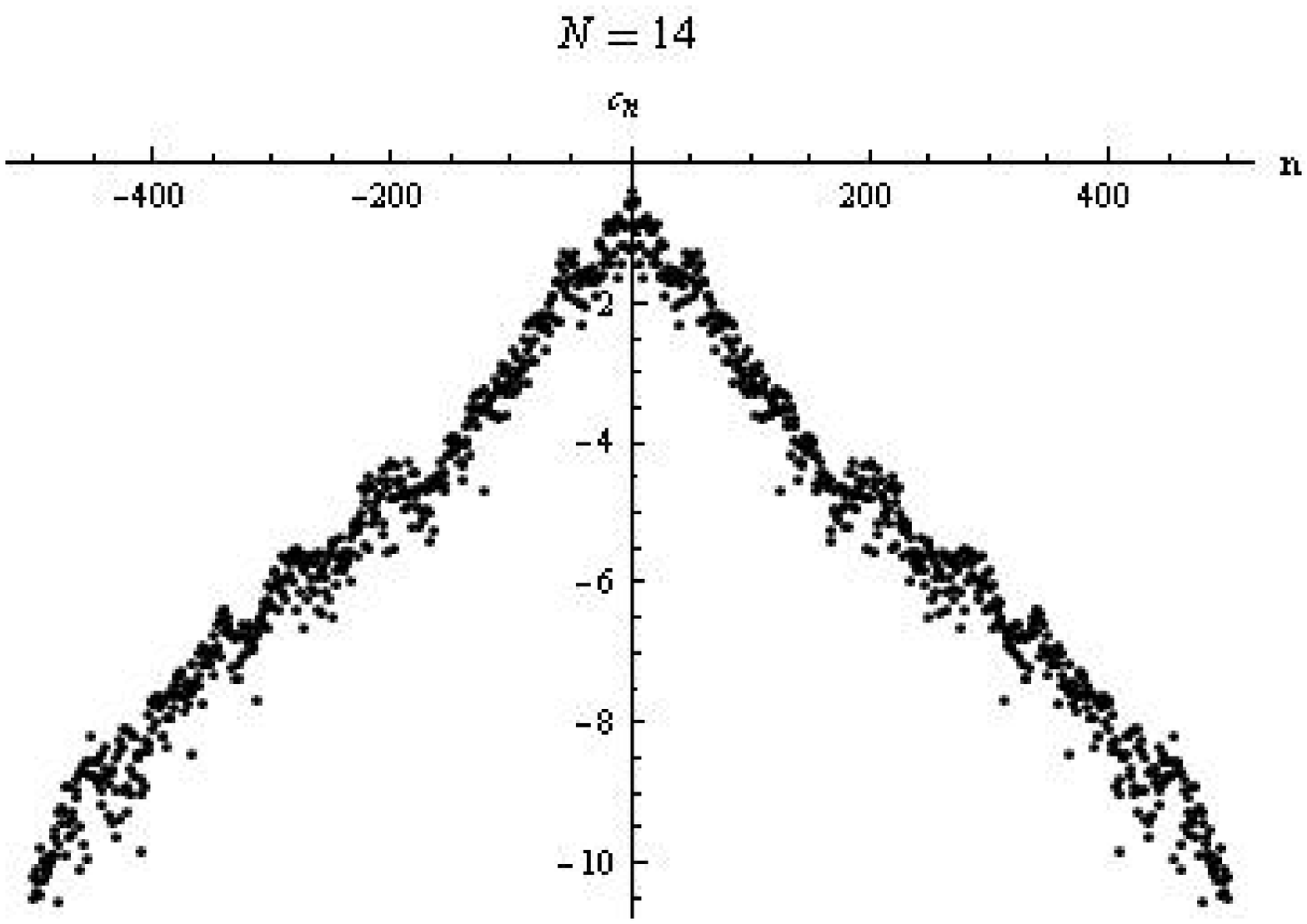}
   \end{minipage}
   \begin{minipage}{17.0 cm}
   \includegraphics[width=5.5cm]{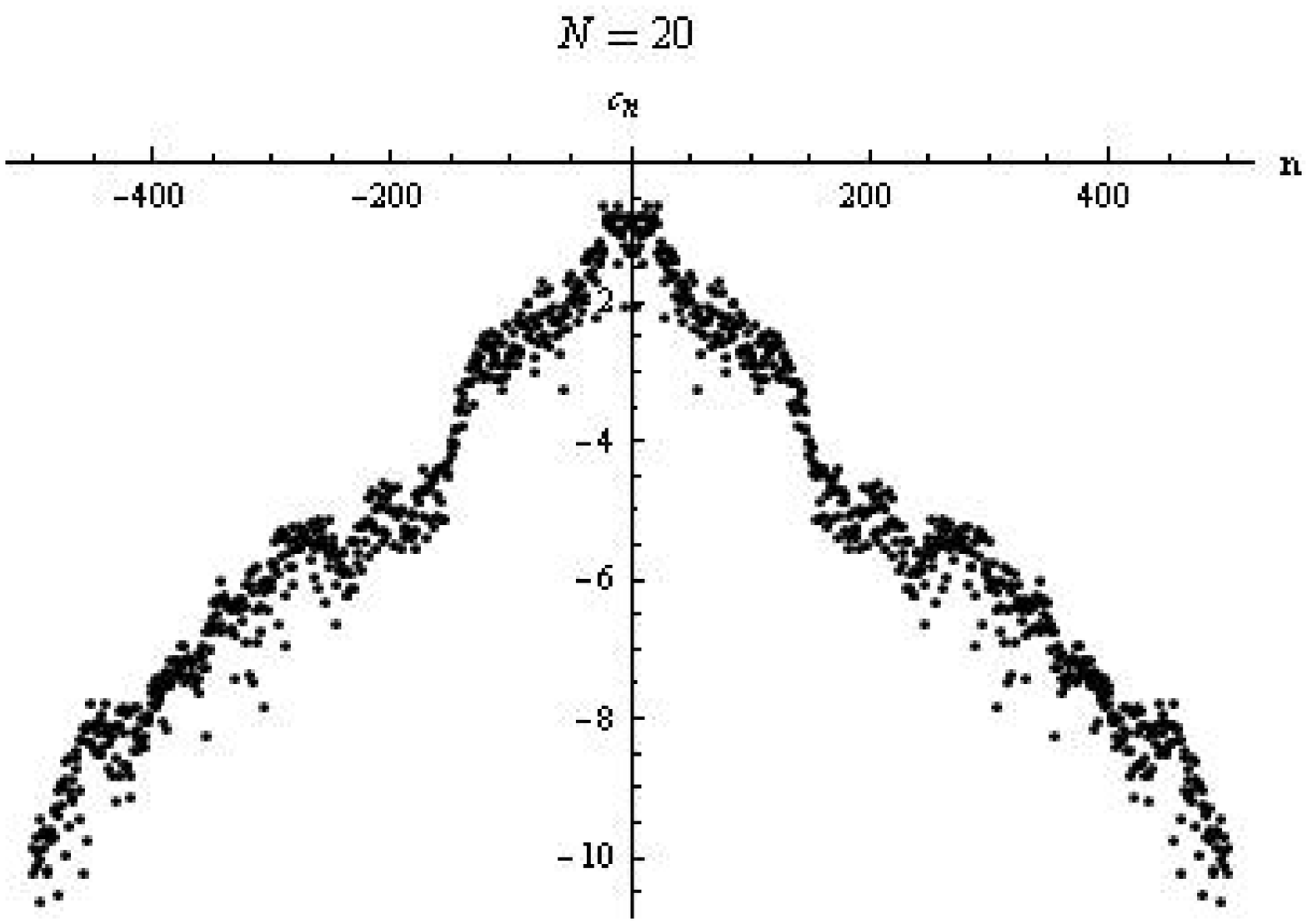}
   \includegraphics[width=5.5cm]{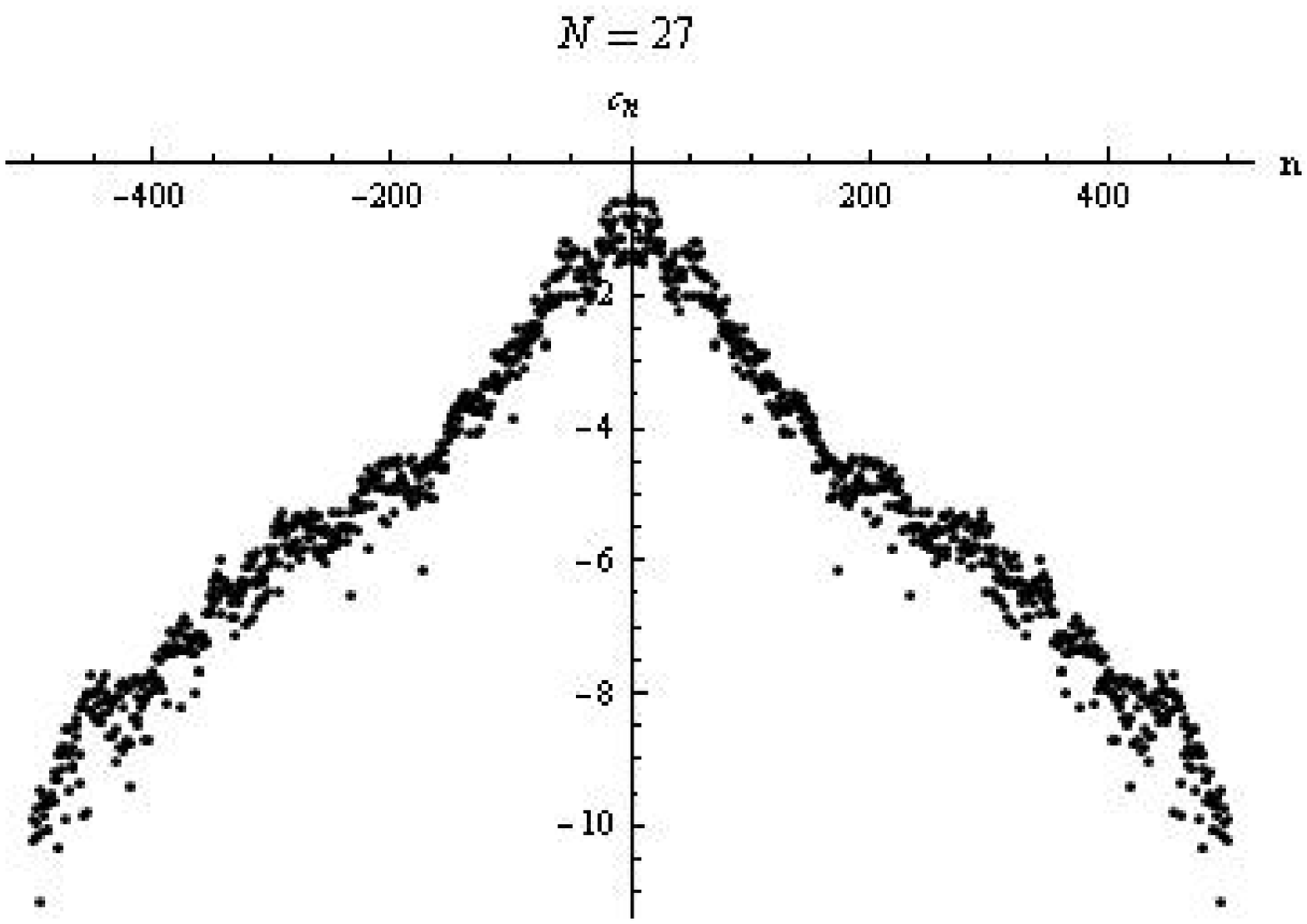}
   \includegraphics[width=5.5cm]{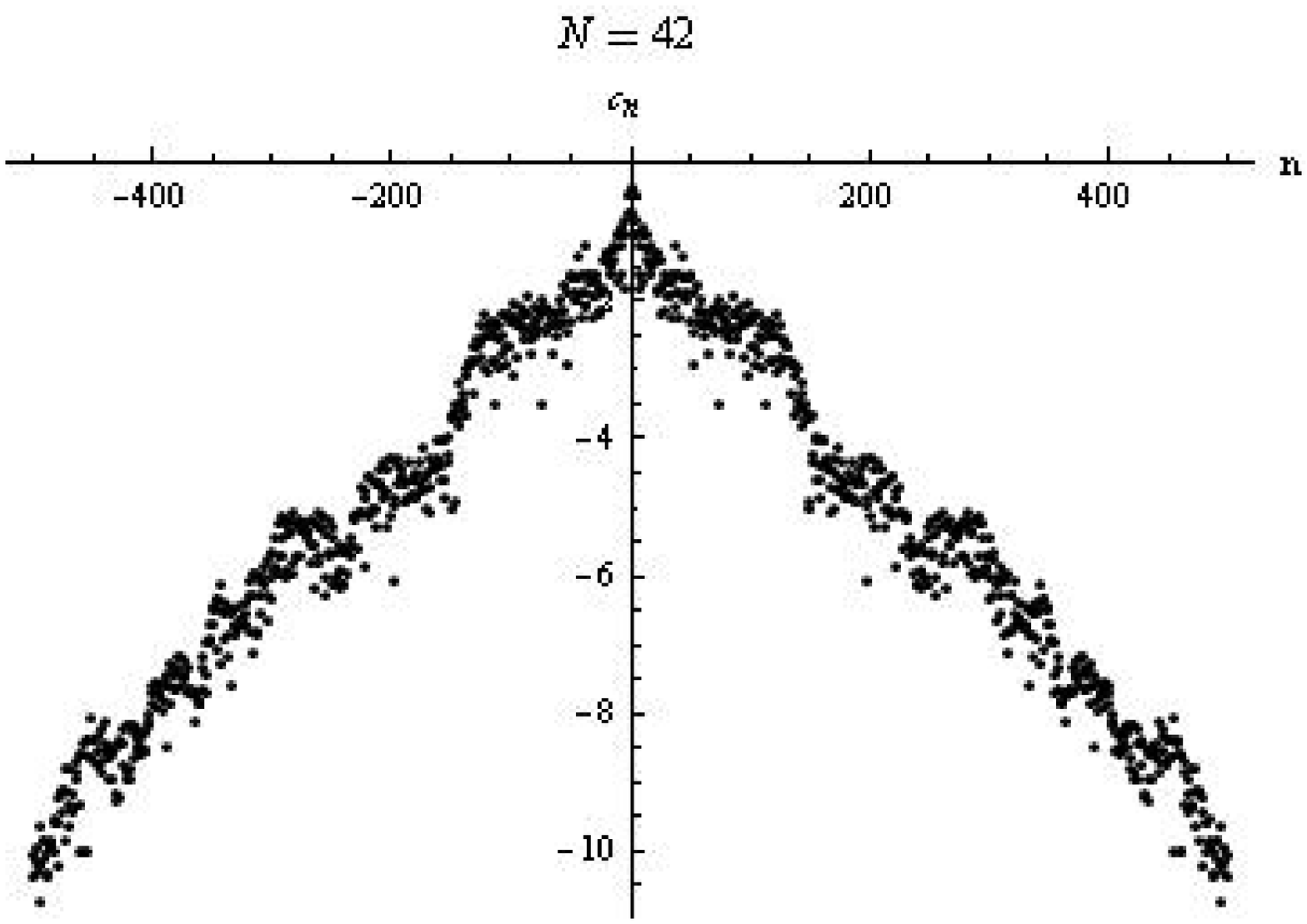}
   \end{minipage}
   \begin{minipage}{17.0 cm}
   \includegraphics[width=5.5cm]{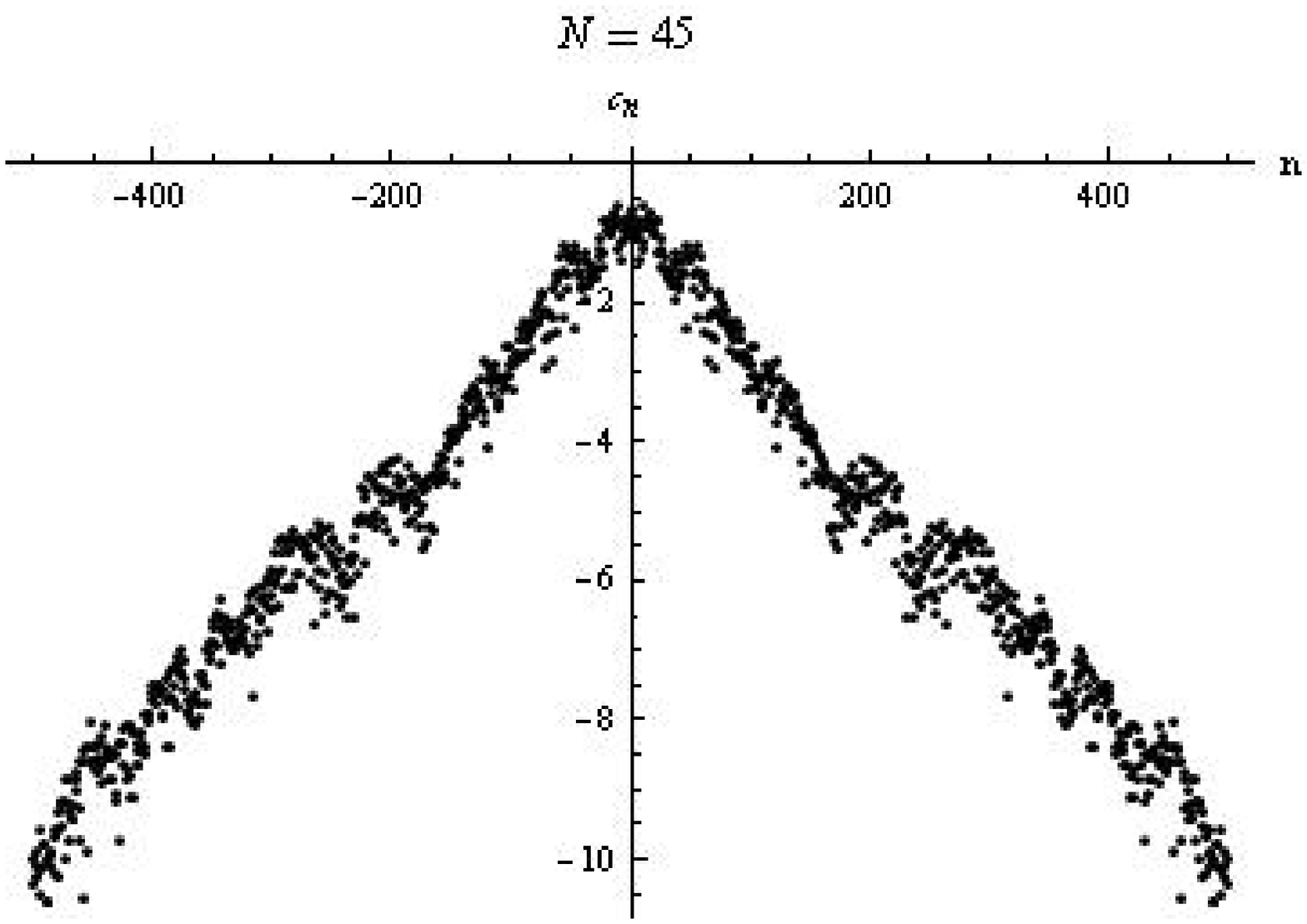}
   \includegraphics[width=5.5cm]{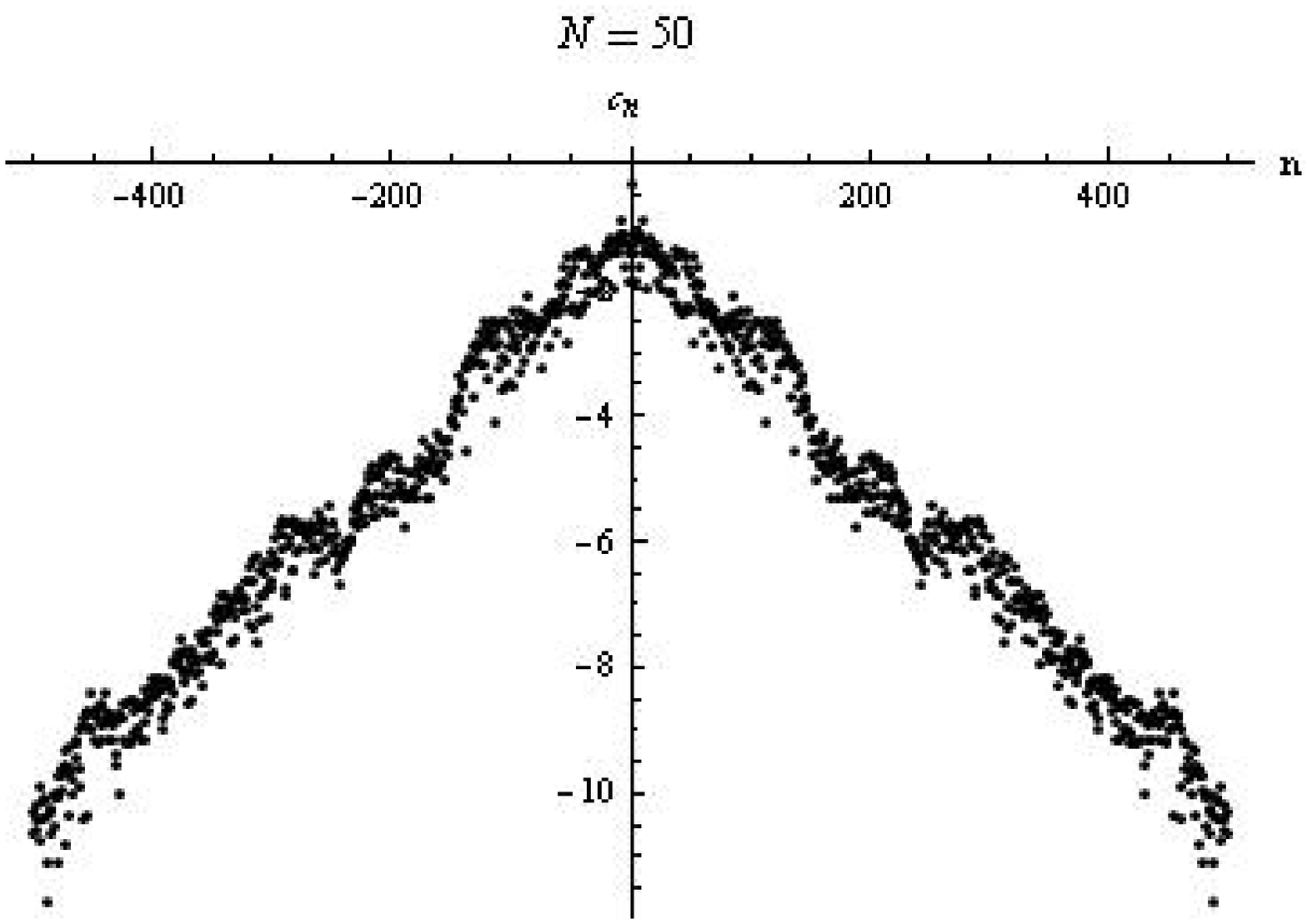}
   \includegraphics[width=5.5cm]{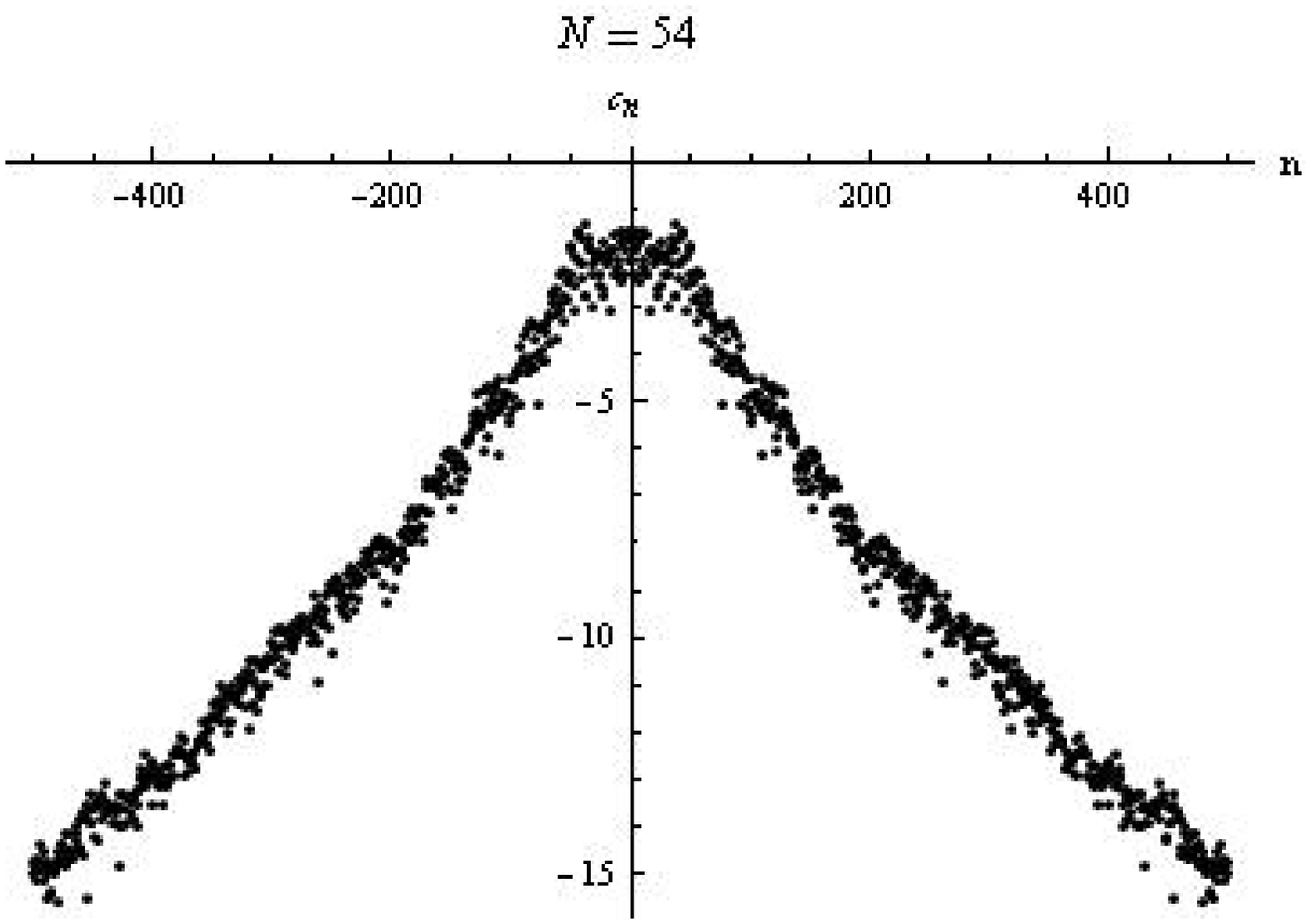}

\caption{QKPR wave function at different time for $k=5$, $\tau=2 \pi
\frac{1}{3}$.}
\end{minipage}
\end{center}
\end{figure*}

\begin{figure*}
\begin{center}
   \begin{minipage}{17.0 cm}
   \includegraphics[width=16cm]{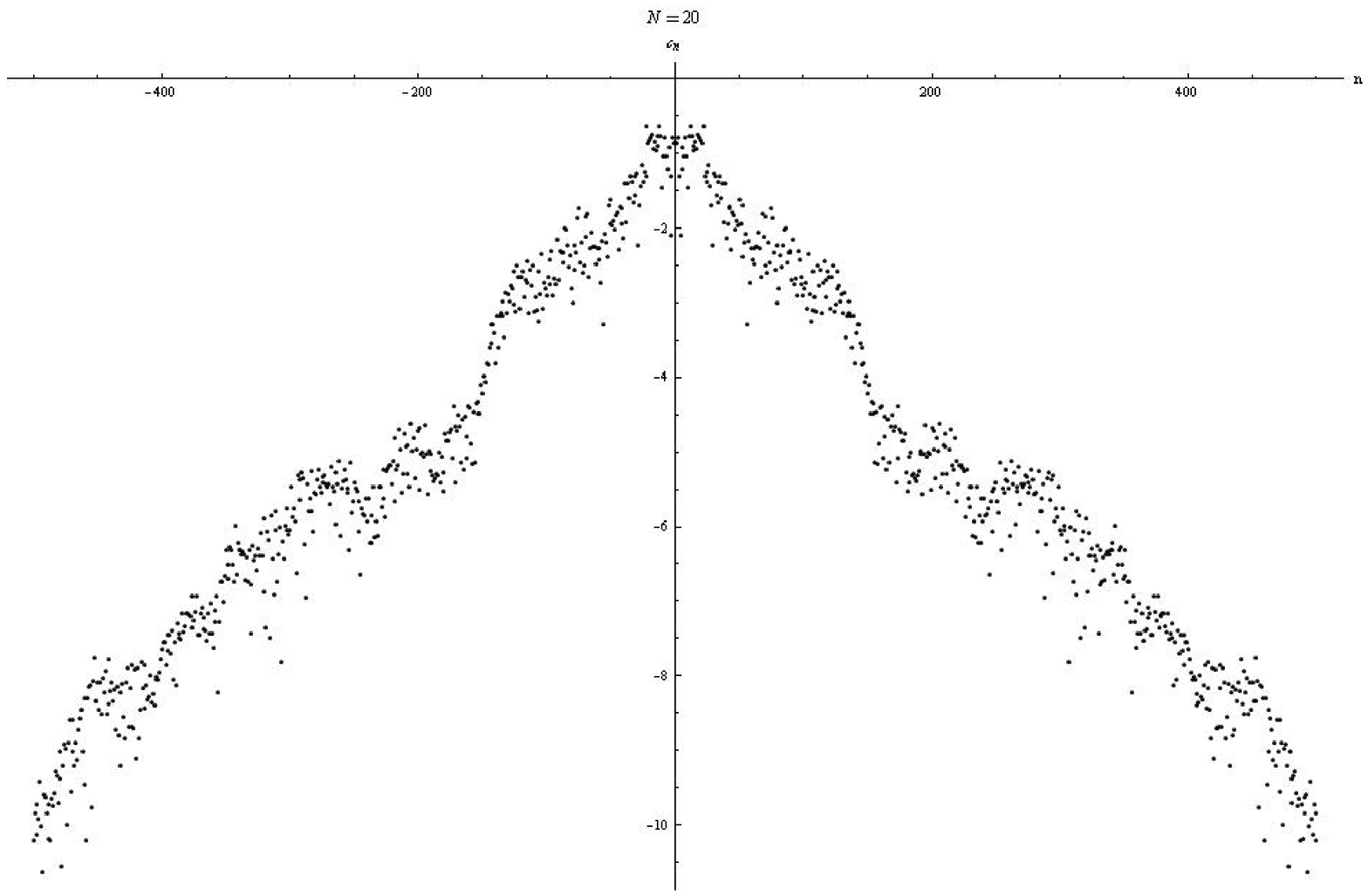}
   \includegraphics[width=16cm]{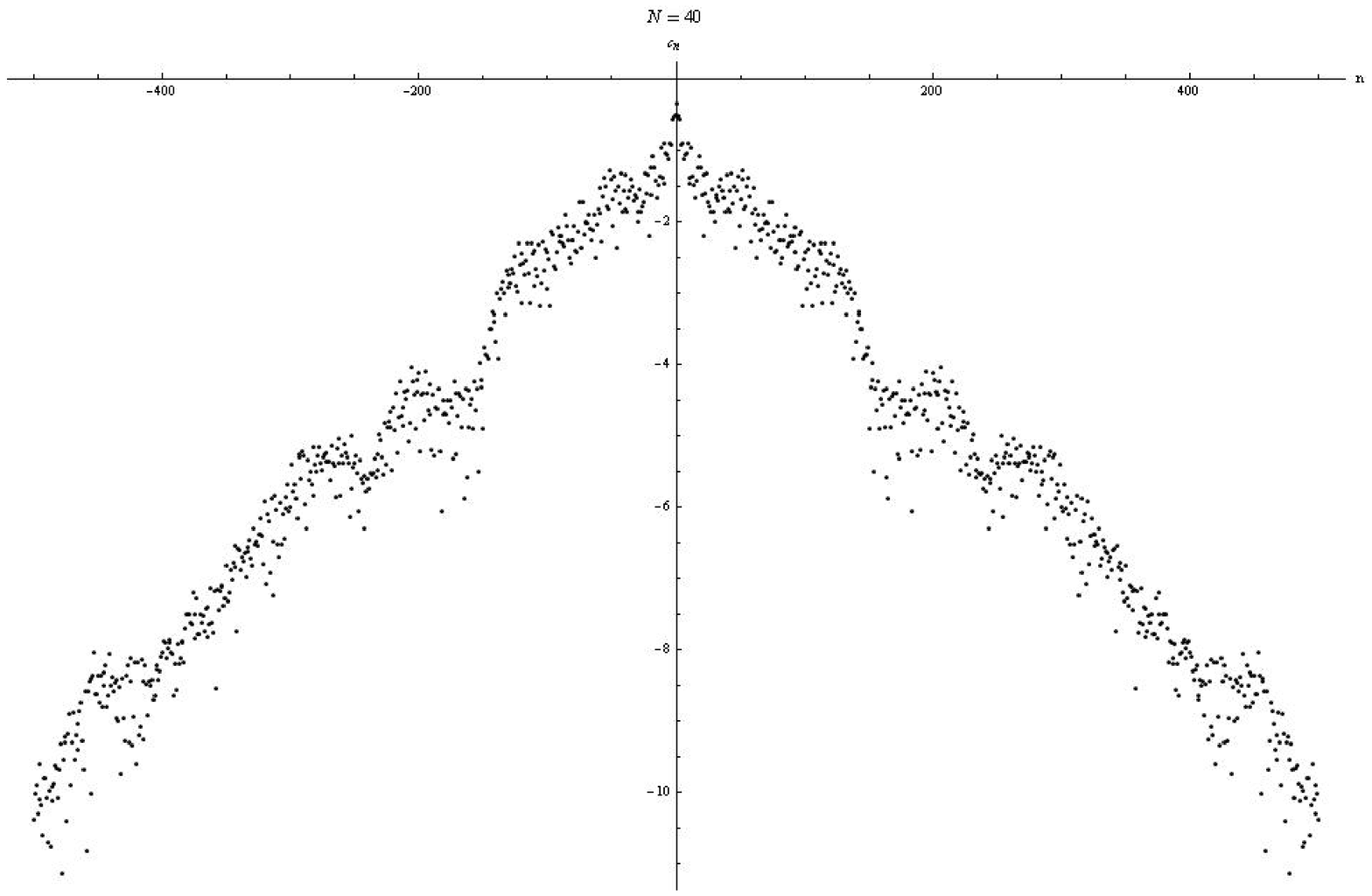}
\caption{QKPR wave function plateaux and cliffs at $N=20$ and $N=40$
for $k=5$, $\tau=2 \pi \frac{1}{3}$.}
\end{minipage}
\end{center}
\end{figure*}


\vspace*{-.5cm}
\begin{thebibliography}{99}
\vspace*{-.5cm}
\bibitem{TaoMa2007} T. Ma, nlin/0709.2395.
\bibitem{Fishman1982} S. Fishman, D.R. Grempel, and R.E. Prange, Phys. Rev. Lett. {\bf 49}, 509
(1982).
\bibitem{Bovier1991} A. Bovier, J. Phys. A Math. Gen. {\bf 25}1021-1029
(1992).
\bibitem{Dunlap1989} D. H. Dunlap, H-L. Wu and P. W. Phillips, Phys.
Rev. Lett. {\bf 65}, 88 (1989).
\bibitem{Chirikov1981} B. V. Chirikov, F. M. Izrailev and D. L. Shepelyansky, Sov. Sci. Rev. C {\bf 2}, 209 (1981).
\bibitem{Chirikov1988} B. V. Chirikov, F. M. Izrailev and D. L. Shepelyansky, Physica D {\bf 33}, 77 (1988).
\bibitem{Shepelyansky1986} D. L. Shepelyansky, Phys. Rev. Lett. {\bf 56}, 677 (1986).
\bibitem{TaoMa2007General}T. Ma, nlin/0709.2735.
\end{thebibliography}
\end{document}